\documentclass[runningheads,a4paper]{llncs}

\usepackage[utf8]{inputenc}

\usepackage{enumitem}

\usepackage{amssymb}
\usepackage{graphicx}
\usepackage{multirow}
\usepackage{ mathrsfs }

\usepackage{listings}

\usepackage{mathtools}

\usepackage[utf8]{inputenc}
\usepackage[english]{babel}
\usepackage{ upgreek }
\usepackage{ amssymb }
\usepackage{ stmaryrd }
\usepackage{graphicx}
\usepackage{amsmath}
\usepackage{amssymb}

\usepackage{fancyhdr}

\usepackage{amsmath}

\usepackage{showexpl}

\usepackage[usenames, dvipsnames]{color} 

\lstdefinestyle{numbers}
{numbers=left, stepnumber=1, numberstyle=\tiny, numbersep=10pt}
\lstdefinestyle{nonumbers}
{numbers=none}

\graphicspath{ {images/} }

\newtheorem{theorem-non}{}

\ifdefined \UseFullVersion 

\else

\fi

\ifdefined \KeepNotesComments 
	\newcommand{\notes}[1]{} 
	\newcommand{\todo}[1]{} 
	\newcommand{\todoh}[1]{} 

	\newcommand{\todohide}[1]{} 
	\newcommand{\notesx}[1]{} 
	\newcommand{\notesy}[1]{} 
	\newcommand{\notesh}[1]{} 
	\newcommand{\removed}[1]{} 
	\newcommand{\done}[1]{} 

	\newcommand{\todon}[1]{\medskip \noindent \textcolor{orange}{NOTES (WILL BE REMOVED LATER): #1} \medskip} 
	
	\newcommand{\notesnta}[1]{\medskip \noindent \textcolor{VioletRed}{NOTES (WILL BE REMOVED LATER): #1} \medskip} 

\else 
	\newcommand{\willdelete}[1]{} 
	\newcommand{\notes}[1]{} 
	\newcommand{\todo}[1]{} 
	\newcommand{\todoh}[1]{} 

	\newcommand{\todohide}[1]{} 
	\newcommand{\notesx}[1]{} 
	\newcommand{\notesy}[1]{} 
	\newcommand{\notesh}[1]{} 
	\newcommand{\removed}[1]{} 
	\newcommand{\done}[1]{} 

	\newcommand{\notesnta}[1]{} 
	\newcommand{\todon}[1]{} 
\fi

	\newcommand{\changenaa}[1]{#1} 
\newcommand{\changenaab}[1]{#1} 
\newcommand{\changenaac}[1]{#1} 

\ifdefined \DifferentColorForTextToModify 

	\newcommand{\change}[1]{#1} 
	\newcommand{\changez}[1]{#1} 
	\newcommand{\changed}[1]{#1} 
	\newcommand{\changex}[1]{#1} 
	\newcommand{\changee}[1]{#1} 
	\newcommand{\changea}[1]{#1} 
	\newcommand{\changec}[1]{#1} 
	\newcommand{\changeb}[1]{#1} 

	\newenvironment{mycolorforparas}{
	\leavevmode\color{orange}\ignorespaces%
	}{%
	}%

	\newenvironment{myColorForRemovePages}{
	
	\leavevmode\color{orange}\ignorespaces%
	}{%
	}%

	\ifdefined \OneColorForTheLastChange 

		\newcommand{\changeh}[1]{\textcolor{blue}{#1}} 

		\newcommand{\changenb}[1]{\textcolor{blue}{#1}} 
		
		\newcommand{\changend}[1]{\textcolor{blue}{#1}} 
		\newcommand{\changene}[1]{\textcolor{blue}{#1}} 

		\newcommand{\changencd}[1]{\textcolor{blue}{#1}} 

		\newcommand{\changenk}[1]{\textcolor{blue}{#1}} 
		\newcommand{\changenr}[1]{\textcolor{blue}{#1}} 

		\newcommand{\changenrlangle}[1]{\textcolor{blue}{#1}} 
		
		\newcommand{\changenta}[1]{\textcolor{blue}{#1}} 

		\newcommand{\changenrp}[1]{\textcolor{blue}{#1}} 
		\newcommand{\changenrpb}[1]{\textcolor{blue}{#1}} 
		\newcommand{\changenrpc}[1]{\textcolor{blue}{#1}} 
		\newcommand{\changenrpd}[1]{\textcolor{blue}{#1}} 

	\else

		\newcommand{\changeh}[1]{\textcolor{red}{#1}} 

		\newcommand{\changenb}[1]{\textcolor{blue}{#1}} 
		
		\newcommand{\changend}[1]{\textcolor{red}{#1}} 
		\newcommand{\changene}[1]{\textcolor{red}{#1}} 

		\newcommand{\changencd}[1]{\textcolor{orange}{#1}} 

		\newcommand{\changenk}[1]{\textcolor{red}{#1}} 
		\newcommand{\changenr}[1]{\textcolor{red}{#1}} 

		\newcommand{\changenrlangle}[1]{\textcolor{blue}{#1}} 
		
		\newcommand{\changenta}[1]{\textcolor{red}{#1}} 

		\newcommand{\changenrp}[1]{\textcolor{orange}{#1}} 
	\newcommand{\changenrpb}[1]{\textcolor{orange}{#1}} 
	\newcommand{\changenrpc}[1]{\textcolor{violet}{#1}} 
		\newcommand{\changenrpd}[1]{\textcolor{violet}{#1}} 

	\fi

\else

	\newenvironment{myColorForRemovePages}{
	
	}{%
	}%

	\newcommand{\change}[1]{#1} 
	\newcommand{\changex}[1]{#1} 
	\newcommand{\changed}[1]{#1} 
	\newcommand{\changez}[1]{#1} 
	\newcommand{\changea}[1]{#1} 
	\newcommand{\changeb}[1]{#1} 
	\newcommand{\changec}[1]{#1} 
	\newcommand{\changee}[1]{#1} 
		\newcommand{\changenr}[1]{#1} 
			\newcommand{\changenk}[1]{#1} 
				\newcommand{\changeh}[1]{#1} 

		\newcommand{\changencd}[1]{#1} 
		\newcommand{\changenrlangle}[1]{#1} 
	\newcommand{\changenrpd}[1]{#1} 
	\newcommand{\changene}[1]{#1} 
	\newcommand{\changenb}[1]{#1} 
	\newcommand{\changend}[1]{#1} 
	\newcommand{\changenta}[1]{#1} 
	\newcommand{\changenrp}[1]{#1} 
	\newcommand{\changenrpb}[1]{#1} 
	\newcommand{\changenrpc}[1]{#1} 
	
\fi

\usepackage[svgnames]{xcolor}
\definecolor{darkgreen}{rgb}{0, 0.5, 0}
\definecolor{darkpurple}{rgb}{0.7, 0, 0.7}
\definecolor{darkblue}{rgb}{0, 0, 0.7}

\usepackage{graphicx}
\usepackage{caption}
\usepackage{subcaption}

\usepackage{tikz}
\usetikzlibrary{shapes,arrows}
\tikzstyle{block} = [draw, fill=blue!20, rectangle, 
minimum height=3em, minimum width=6em]
\tikzstyle{sum} = [draw, fill=blue!20, circle, node distance=1cm]
\tikzstyle{input} = [coordinate]
\tikzstyle{output} = [coordinate]
\tikzstyle{pinstyle} = [pin edge={to-,thin,black}]

\usepackage{amsthm}
\newtheoremstyle{exampstyle}
{0.3} 
{0.3} 
{} 
{} 
{\bfseries} 
{.} 
{.1em} 
{} 
\theoremstyle{exampstyle} \newtheorem{defn}{Definition}
\theoremstyle{exampstyle} \newtheorem{prop}{Proposition}
\theoremstyle{exampstyle} \newtheorem{mylemma}{Lemma}

\usepackage[
colorlinks=true,
\ifdefined\VersionWithComments
pagebackref=true,
\fi
citecolor=darkgreen,
linkcolor=darkblue,
urlcolor=darkpurple,
]{hyperref}

\ifdefined \VersionWithComments
\usepackage{marginnote}
\newcommand{\marginX}{\marginnote{\huge{\quad\quad\textbf{!}\quad\quad}}}
\newcommand{\ea}[1]{{\color{violet}\marginX{}\textbf{[\'Etienne}: #1]}}
\newcommand{\hv}[1]{{\color{green!50!black}\marginX{}\textbf{[Huu-Vu}: #1]}}
\newcommand{\lp}[1]{{\color{magenta}\marginX{}\textbf{[Laure}: #1]}}
\else
\newcommand{\ea}[1]{}
\newcommand{\hv}[1]{}
\newcommand{\lp}[1]{}
\fi

\begin{document}

\mainmatter 

\title{CARET analysis of multithreaded programs\thanks{\changenaac{This work was partially funded by the FUI project AiC.}}}


\addtolength{\textheight}{1.4cm}
\addtolength{\textwidth}{1cm}  

\titlerunning{CARET analysis of multithreaded programs}

%
%
\author{Huu-Vu Nguyen$^1$, Tayssir Touili$^2$}
\authorrunning{}

\institute{$^1$ University Paris Diderot and LIPN, France \\ $^2$ \changenta{CNRS, LIPN  and University Paris 13, France}}

%
%

\maketitle

\ifdefined \VersionWithComments
\textcolor{red}{\textbf{This is the version with comments. To disable comments, comment out line~3 in the \LaTeX{} source.}}
\fi

\lp{
	Some advice on writing tool papers can be found here: \url{http://www.informatik.uni-hamburg.de/TGI/PetriNets/sc-info/docs/ToolFormat.pdf}}

\ea{Guidelines: 
	
	\textbf{
		Tools papers of a maximum of 4 pages should describe an operational tool and its contributions; 2 additional pages of appendices are allowed that will not be included in the proceedings. Tool papers should explain enhancements made compared to previously published work. A tool paper need not present the theory behind the tool but can focus more on its features, and how it is used, with screen shots and examples. Authors of tools papers should make their tool available for use by reviewers.
	}
}

\begin{abstract}
	
	Dynamic Pushdown Networks (DPNs) are a natural model for multithreaded programs with (recursive) procedure calls and thread creation. On the other hand, CARET is a temporal logic that allows to write linear temporal formulas while taking into account the matching between calls and returns. We consider in this paper the model-checking
	problem of DPNs against CARET formulas. We show that this problem
	can be effectively solved by a reduction to the emptiness problem of B\"{u}chi
	Dynamic Pushdown Systems. We then show that CARET model checking
	is also decidable for DPNs communicating with locks. Our results
	can, in particular, be used for the detection of concurrent malware.
\end{abstract}


\todohide{Changes Locks part - change create D in intuition-- remove langle ,rangle in Proofs}

\notesnta{In the conference page, thay said that we need to add email to the author information? WHAT DO YOU THINK? I WILL ADD OR NOT?}

\section{Introduction}

\change{Pushdown Systems (PDSs) are known to be a natural model for sequential programs \cite{schwoonThesis}. Therefore, networks of pushdown systems are a natural model for concurrent programs where each PDS represents a sequential \changenta{component} of the system. In this context, Dynamic pushdown Networks (DPNs) \cite{DBLP:conf/concur/BouajjaniMT05} were introduced by Bouajjani et al. as a natural model of multithreaded programs with  procedure calls and thread creation.} Intuitively, a DPN is a network of pushdown processes $\{ \mathcal{P}_1, ..., \mathcal{P}_n \}$ where each process, represented by a Pushdown system (PDS), can perform basic pushdown actions, call procedures, as well as spawn \changez{new instances of pushdown processes}. A lot of previous researches focused on investigating automated methods to verify DPNs. In \cite{DBLP:conf/concur/BouajjaniMT05,DBLP:journals/ijfcs/Lugiez11,DBLP:conf/cav/LammichMW09,DBLP:conf/vmcai/GawlitzaLMSW11}, the reachability analysis of DPNs are considered. While the model-checking problem for DPNs against double-indexed properties is undecidable, \change{i.e.}, the properties where the satisfiability of an atomic proposition depends on control states of two or more threads \cite{DBLP:conf/lics/KahlonG06}, it is \changez{decidable to model-check DPNs against the linear temporal logic} (LTL) and the computation tree logic (CTL) with single-indexed properties \cite{DBLP:conf/aplas/SongT13}, \change{i.e.}, \changez{properties} where the satisfiability of an atomic proposition depends on control states of only one thread.

\done{Motivation for the need of CARET in concurrent program. How to say well here???}
%

CARET is a temporal logic of calls and returns \cite{DBLP:conf/tacas/AlurEM04}. This logic allows us to write linear temporal formulas while taking into account the matching between calls and returns. \changez{CARET is needed to describe several important properties such as malicious behaviors or API usage rules. Thus, to be able to analyse such properties for multithreaded programs, we need to be able to check CARET formulas for DPNs. We tackle this problem in this paper. As LTL is a subclass of CARET, CARET model-checking for DPNs with double-indexed properties is also undecidable. Thus, in this paper, we consider the model-checking problem} for DPNs against single-indexed CARET formulas and show that it is decidable. A single-indexed CARET formula is a formula in the form $\bigwedge f_i$ where $f_i$ is a CARET formula over a certain PDS $\mathcal{P}_i$. A DPN satisfies $\bigwedge f_i$ iff all instances of the PDS \changenr{$\mathcal{P}_i$} created in the network \changenta{satisfy} the subformula $f_i$. 

The model-checking problem of DPNs against single-indexed CARET formulas is non-trivial because the number of instances of pushdown processes in DPNs can be unbounded. \change{It is not sufficient to check if every PDS $\mathcal{P}_i$ satisfies the corresponding formula $f_i$. Indeed, we need to ensure that all instances  of $\mathcal{P}_i$ created during a run of DPN satisfies the formula $f_i$. Also, it is not correct to check whether all possible instances of $\mathcal{P}_i$ \changenta{satisfy} the formula $f_i$. Indeed, an instance of $\mathcal{P}_i$ should not be checked if it is not created during the run of DPNs. In this paper, we solve these problems. \changez{We show that \changenta{single-indexed} CARET model checking is decidable for DPNs. To this end, we reduce the problem of checking whether} Dynamic Pushdown Networks \changenta{satisfy} single-indexed CARET formulas to the membership problem for B\"{u}chi Dynamic Pushdown Networks (BDPNs). \changez{Finally, we show that single-indexed CARET  model checking is decidable for Dynamic Pushdown Networks communicating via nested locks.}}

\medskip
\noindent
\textbf{Related work}. 




\notes{I'm not sure what is original paper about Multi PDS}


\cite{DBLP:conf/popl/BouajjaniET03,DBLP:conf/tacas/ChakiCKRT06,DBLP:conf/concur/AtigBT08,DBLP:conf/wia/AtigT09} considered Pushdown networks with communications between processes. 	\changez{However, these works consider only  networks with a fixed number of threads. The model-checking problem for pushdown networks where synchronization between threads is ensured by a set of nested locks is considered in \cite{DBLP:conf/cav/KahlonIG05,DBLP:conf/lics/KahlonG06,DBLP:conf/popl/KahlonG07} for single-indexed LTL/CTL and double-indexed LTL. These works do not handle dynamic thread creation.} 
 
\changed{Multi-pushdown systems were considered in \cite{DBLP:conf/ifipTCS/TorreN12,DBLP:conf/csr/BansalD13} to represent \changenaac{multithreaded} programs. These systems have only a finite number of stacks, and thus, they
	cannot handle dynamic thread creation.}

\done{Multi-pushdown systems were considered in [] to represent multitheraded programs. These systems have only a finite number of stacks, and thus, they cannot handle dynamic therad creation. TALK ABOUT CARET FOR MPDSS HERE!!}


Pushdown Networks with dynamic thread creation (DPNs) \changez{were} introduced in \cite{DBLP:conf/concur/BouajjaniMT05}. The reachability problems of DPNs and its extensions are considered in  \cite{DBLP:conf/concur/BouajjaniMT05,DBLP:conf/vmcai/GawlitzaLMSW11,DBLP:conf/cav/LammichMW09,DBLP:journals/ijfcs/Lugiez11,DBLP:conf/esop/Wenner10}. \cite{DBLP:conf/aplas/SongT13} considers the model-checking problem of DPNs against single-indexed \changez{LTL and CTL,} while \cite{DBLP:journals/corr/SongT16} investigates the single-indexed LTL \changez{model checking problem for DPNs with locks.}

\changenrpb{\cite{DBLP:conf/SAC/VuTayssir2017,DBLP:conf/SPIN/VuTayssir2017} consider CARET model checking for pushdown systems and its application to malware detection. These works can only handle sequential programs. In this paper, we go one step further and extend these works \cite{DBLP:conf/SAC/VuTayssir2017,DBLP:conf/SPIN/VuTayssir2017} to DPNs and concurrent programs.}


\done{OK. will do it -> Maybe should add comparision with Multi-CARET for multi-PDS here}



%
%

\section{Linear Temporal Logic of Calls and Returns - CARET}

\label{sec:CARETDefineAtomDefine}


In this section, we recall the definition of CARET \cite{DBLP:conf/tacas/AlurEM04}. A CARET formula is interpreted on an infinite path where each state on the path is associated with a tag in the set $\{call, ret, int \}$. A \textit{call-state} denotes an invocation to a procedure of a program while the corresponding \textit{ret-state} denotes the \textit{ret} statement of that procedure. A \textit{simple} statement (neither a \textit{call} nor a \textit{ret} statement) is called an \textit{internal} statement and its associated state is called \textit{int-state}.

Let $\omega = s_0 s_1 ...$ be an infinite path where each state on the path is associated with a tag in the set $\{ call, ret, int \}$. Over $\omega$, three kinds of successors are defined for every position $s_i$:
\ifdefined \NotNeedToReducePages \begin{itemize} \else \begin{itemize}[noitemsep,topsep=0pt] \fi
		\item{\textit{global-successor}: The global-successor of $s_i$ is $s_{i+1}$.}
		\item{\textit{abstract-successor}: The abstract-successor of $s_i$ is determined by its associated tag.
			\ifdefined \NotNeedToReducePages \begin{itemize} \else \begin{itemize}[noitemsep,topsep=0pt] \fi
					\item{If $s_i$ is a \textit{call}, the abstract successor of $s_i$ is the matching return point.}
					\item{\changex{If $s_i$ is a \textit{int}, the abstract successor of $s_i$ is $s_{i+1}$.}}
					
					\item{\changex{If $s_i$ is a \textit{ret}, the abstract successor of $s_i$ is defined as $\bot$.}}

				\end{itemize}
			}
			\item{\textit{caller-successor}: The caller-successor of $s_i$ is the most inner unmatched call if there is such a \textit{call}. Otherwise, it is defined as $\bot$.}
		\end{itemize}
		
		\ifdefined \NotNeedToReducePages	
		
		\begin{myColorForRemovePages}

			For example, in Figure \ref{fig:terminology}:
			\ifdefined \NotNeedToReducePages \begin{itemize} \else \begin{itemize}[noitemsep,topsep=0pt] \fi
					\item{The global-successor of $s_1$ and $s_2$ are $s_2$ and $s_3$ respectively.}
					\item{The abstract-successor of $s_2$ and $s_5$ are $s_k$ and $s_9$ respectively.}
					\item{The caller-successor of $s_6$, $s_7$, $s_8$ is $s_5$ while the caller-successor of $s_3$, $s_4$, $s_5$, $s_9$ is $s_2$. \change{Note that the caller-successor of $s_0$, $s_1$, $s_2$, $s_k$ is $\bot$}.}
				\end{itemize}
				
			\end{myColorForRemovePages}
			\fi

			\ifdefined \NotNeedToReducePages	
			
			\begin{myColorForRemovePages}
				
				A \textit{global-path} is obtained by applying repeatedly the global-successor operator. Similarly, an \textit{abstract-path} or a \textit{caller-path} are obtained by repeatedly applying the abstract-successor and caller-successor respectively.

			\end{myColorForRemovePages}

			\else
			A \textit{global-path} is obtained by applying repeatedly the global-successor operator. Similarly, an \textit{abstract-path} or a \textit{caller-path} are obtained by repeatedly applying the abstract-successor and caller-successor respectively. 
			
			
			\fi

			%
			

			\ifdefined \NotNeedToReducePages	
			
			\begin{myColorForRemovePages}
				
				In Figure \ref{fig:terminology}, from $s_4$, the global-path is $s_4 s_5 s_6 s_7 s_8 s_9 s_{10} ...$, the abstract-path is $ s_4 s_5 s_9 s_{10} ...$ while the caller-path is $s_4 s_2$. Note that the caller-path is always finite.
				\begin{figure*}
					\centering
					\begin{tikzpicture}[xscale=1, yscale=1.3]
					\tikzset{lineStyle/.style={blue, thick, ->}};
					\tikzset{myDot/.style={blue, fill = blue, thick}};
					\tikzset{myRectangleNode/.style={rectangle, thick, draw= black, below right, black}};
					\draw [lineStyle] (0, 0) -- (1,0);
					\draw [lineStyle] (1,0) -- (2,0) ;
					\draw [lineStyle] (2,0) -- (2.5, -1);
					\draw [lineStyle] (2.5, -1) -- (3.5, -1);
					\draw [lineStyle] (3.5, -1) -- (4.5, -1);
					\draw [lineStyle] (4.5, -1) -- (5, -2) ;
					\draw [lineStyle] (5, -2) -- (6, -2);
					\draw [lineStyle] (6, -2) -- (7, -2) ;
					\draw [lineStyle] (7, -2) -- (7.5, -1) -- (8.5, -1);
					\draw [dotted] (8.5, -1) -- (9.5, -1);
					\draw [lineStyle] (10.5, -1) -- (11, 0);
					\draw [dotted] (11, 0) -- (12, 0);
					\draw [myDot] (0,0) circle [radius=0.04];
					\draw [myDot] (1,0) circle [radius=0.04];
					\draw [myDot] (2,0) circle [radius=0.04];
					\draw [myDot] (2.5, -1) circle [radius=0.04];
					\draw [myDot] (3.5, -1) circle [radius=0.04];
					\draw [myDot] (4.5, -1) circle [radius=0.04];
					\draw [myDot] (5, -2) circle [radius=0.04];
					\draw [myDot] (6, -2) circle [radius=0.04];
					\draw [myDot] (7, -2) circle [radius=0.04];
					\draw [myDot] (7.5, -1) circle [radius=0.04];
					\draw [myDot] (8.5, -1) circle [radius=0.04];
					\draw [myDot] (10.5, -1) circle [radius=0.04];
					\draw [myDot] (11, 0) circle [radius=0.04];
					\draw [myDot] (12, 0) circle [radius=0.04];
					\node[below left]() at (0,0) {$s_0$};
					\node[below left]() at (1,0) {$s_1$};
					\node[below left]() at (2,0) {$s_2$};
					\node[below left]() at (2.5,-1) {$s_3$};
					
					\node[ left]() at (2,-1) {\changeh{$proc$}};

					\node[below left]() at (3.5, -1) {$s_4$};
					\node[below left]() at (4.5, -1) {$s_5$};
					\node[below left]() at (5, -2) {$s_6$};
					\node[below left]() at (6, -2) {$s_7$};
					\node[below left]() at (7, -2) {$s_8$};
					\node[below left]() at (7.5, -1) {$s_9$};
					\node[below left]() at (8.5, -1) {$s_{10}$};
					\node[below left]() at (11,0){$s_{k}$};
					\node[above]() at (2,0) {$call$};
					\node[above]() at (4.5, -1) {$call$};
					\node[above left]() at (7, -2) {$ret$};
					
					\node[above left](node1) at (2,0) {};
					\node[above left](node1) at (11,0) {};
					\draw[dotted, red, thick, ->] (2,0) .. controls(6,0.5) .. (11,0);
					\draw[dotted, red, thick, ->] (4.5,-1) .. controls(5.5, -0.8) .. (7.5, -1);
					\draw [->, loosely dashdotted] (4.5,-1) -- (2.1, -0.07) ;
					\draw [->, loosely dashdotted] (8.5,-1) -- (2.15, -0.04) ;
					\node[right]() at (1, -1.5) {global-successor};
					\draw[lineStyle] (0, -1.5) -- (1, -1.5);
					\node[right]() at (1, -2) {abstract-successor};
					\draw[->, dotted, red, thick] (0, -2) -- (1, -2);
					\node[right]() at (9, -2) {caller-successor};
					\draw[->, loosely dashdotted] (8, -2) -- (9, -2);
					\end{tikzpicture}
					\caption{Three kinds of successors of CARET} \label{fig:terminology}
				\end{figure*}

				\notesh{I just put back the Figure and example because In iFM paper, one author doesn't understand what is caller successor. I THINK THAT MAYBE IT IS BETTER TO use $p_i \omega_i$ directly to be easy define $\changenrlangle{p \omega }$. }

			\end{myColorForRemovePages}
			\fi

			\ifdefined \NotNeedToReducePages	
			\noindent
			\changenrp{Given a state $s$, let $\mathscr{P}(s)$ be the procedure to which $s$ belongs. For example, in Figure \ref{fig:terminology}, all configurations on the abstract-path starting from $s_3$ \changeh{belong} to the procedure $proc$, i.e., $\mathscr{P}(s_3)$ = $proc$, $\mathscr{P}(s_9)$ = $proc$, ...}
			
			\else
			
			\fi

			\medskip
			\noindent
			\textbf{Formal Definition}. Given a finite set of atomic propositions \textit{AP}. Let $AP' = AP \cup \{ call, ret, int \} $. A CARET formula over AP is defined as follows (where $e \in AP'$):
			$$\psi := e \;|\; \psi \vee \psi \;|\; \neg \psi \;|\; X^{g} \psi \;|\; X^{a} \psi \;|\; X^{c} \psi \;|\; \psi U^{a} \psi \;|\; \psi U^{g} \psi \;|\; \psi U^{c} \psi$$
			
			Let $\Sigma = 2^{AP} \times \{call, ret, int\}$. Let $\uppi=\uppi(0)\uppi(1)\uppi(2)...$ be an $\omega$-word over $\Sigma$. Let $(\uppi, i)$ be the suffix of $\uppi$ starting from $\uppi(i)$. Let $next^{g}_i$, $next^{a}_i$, $next^{c}_i$ be the global-successor, abstract-successor and caller-successor of $\uppi(i)$ respectively.
			The satisfiability relation is defined inductively as follows:
			\ifdefined \NotNeedToReducePages \begin{itemize} \else \begin{itemize}[noitemsep,topsep=0pt] \fi
					\item{$ (\uppi, i) \vDash e$, where $e \in AP'$, iff $\uppi (0)=(Y,d)$ and $e \in Y$ or $e=d$}
					\item{$ (\uppi, i) \vDash \psi_{1} \vee \psi_{2}$ iff $ (\uppi, i) \vDash \psi_{1} $ or $ (\uppi, i) \vDash \psi_{2}$}
					\item{$ (\uppi, i) \vDash \neg \psi$ iff $ (\uppi, i) \nvDash \psi $ }
					\item{$ (\uppi, i) \vDash X^{g} \psi $ iff $(\uppi, next^{g}_i) \vDash \psi$ }
					\item{$ (\uppi, i) \vDash X^{a} \psi $ iff $ next^{a}_i \neq \bot $ and $(\uppi, next^{a}_i) \vDash \psi$ }
					\item{$ (\uppi, i) \vDash X^{c} \psi $ iff $ next^{c}_i \neq \bot $ and $(\uppi, next^{c}_i) \vDash \psi$ }
					\item{$ (\uppi, i) \vDash \psi_{1} U^{b} \psi_{2}$ (with $b\in \{g,a,c\}$) iff there exists a sequence of positions $h_{0},h_{1}, ..., h_{k - 1}, h_{k} $ where $h_{0} = i$, for every $0 \leq j \leq k - 1 $ : $h_{j+1} = next^{b}_{h_{j}}, (\uppi, h_{j}) \vDash \psi_{1}$ and $(\uppi, h_{k}) \vDash \psi_{2}$}
				\end{itemize}

				\ifdefined \NotNeedToReducePages	
				
				\begin{myColorForRemovePages}
					\noindent
					Then, $\uppi \vDash \psi$ iff $(\uppi, 0) \vDash \psi$. Other CARET operators can be expressed by the above operators: $F^g \psi = \text{true }U^g \psi$, $G^g \psi = \neg (\text{true }U^g \neg \psi)$, $F^a \psi = \text{true }U^a \psi$, $G^a \psi = \neg (\text{true }U^a \neg \psi)$, $F^c \psi = \text{true }U^c \psi$, $G^c \psi = \neg \allowbreak (\text{true } \allowbreak U^c \allowbreak \neg \allowbreak \psi)$.

					\notes{I think we can remove this Remark to save space}
					\begin{remark}
						\change{LTL can be seen as the  subclass of CARET
							where the operators $X^c, U^c, X^a, U^a$ are not considered.}
					\end{remark}
					
				\end{myColorForRemovePages}

				\else
				\noindent
						\changenaa{	Then, $\uppi \vDash \psi$ iff $(\uppi, 0) \vDash \psi$. Other CARET operators can be expressed by the above operators: $F^g \psi = \text{true }U^g \psi$, $G^g \psi = \neg (\text{true }U^g \neg \psi)$, $F^a \psi = \text{true }U^a \psi$,...}
				\fi

				\medskip
				\noindent
				\textbf{Closure}. Let $\psi$ be a CARET formula over $AP$. The closure of $\psi$, denoted $Cl(\psi)$, is the smallest set that contains \textbf{$\psi$}, $call$, $ret$ and $int$ and satisfies the following properties:
				\ifdefined \NotNeedToReducePages \begin{itemize} \else \begin{itemize}[noitemsep,topsep=0pt] \fi
						\item{if $\neg \psi' \in Cl(\psi)$, then $\psi' \in Cl(\psi)$}
						\item{if $X^{b} \psi' \in Cl(\psi)$ (with $b \in \{g,a,c\}$), then $\psi' \in Cl(\psi)$}
						\item{if $\psi_{1} \vee \psi_{2} \in Cl(\psi)$, then $\psi_{1} \in Cl(\psi), \psi_{2} \in Cl(\psi)$ }
						\item{if $\psi_{1} U^{b} \psi_{2} \in Cl(\psi)$ (with $b \in \{g,a,c\}$), then $\psi_{1} \in Cl(\psi), \psi_{2} \in Cl(\psi), X^{b} (\psi_{1} U^{b} \psi_{2}) \in Cl(\psi)$}
						\item{if $\psi' \in Cl(\psi)$, and $\psi'$ is not in the form $\neg \psi''$ then $\neg \psi' \in Cl(\psi)$ }
					\end{itemize}
					
					\noindent
					\change{\textbf{Atoms}}. A set $A \subseteq Cl(\psi)$ is an atom of $\psi$ if it satisfies the following properties:
					\ifdefined \NotNeedToReducePages \begin{itemize} \else \begin{itemize}[noitemsep,topsep=0pt] \fi
							\item{$\forall \psi' \in Cl(\psi), \psi' \in A \Leftrightarrow \neg \psi' \notin A$}
							\item{$\forall \psi' \vee \psi'' \in Cl(\psi), \psi' \vee \psi'' \in A \Leftrightarrow \psi' \in A$ or $\psi'' \in A$}
							\item{$\forall \psi' U^{b} \psi'' \in Cl(\psi),$ where $b \in \{ g, a, c \}, \psi' U^{b} \psi'' \in A \Leftrightarrow \psi'' \in A$ or $(\psi' \in A$ and $X^{b}(\psi' U^{b} \psi'') \in A) $}
							
							\item{A includes exactly one element of the set \{call, ret, int\}}
						\end{itemize}
						
						\noindent
						Let $Atoms(\psi)$ be the set of atoms of $\psi$. Let $A$ and $A'$ be two atoms, we define the following predicates:
						\ifdefined \NotNeedToReducePages \begin{itemize} \else \begin{itemize}[noitemsep,topsep=0pt] \fi
								\item{$AbsNext (A, A') = true$ iff for every $X^a \phi' \in Cl(\psi): (X^{a} \phi' \in A$ iff $\phi' \in A')$. }
								\item{ $GlNext (A, A') = true$ iff for every $X^g \phi' \in Cl(\psi): (X^{g} \phi' \in A$ iff $\phi' \in A')$}
								\item{$CallerNext(A, A') = true$ iff for every $X^{c} \phi' \in Cl(\psi): (X^{c} \phi' \in A$ iff $\phi' \in A')$.}
							\end{itemize}

							\changeh{We define $NexCallerForms(A)$ (resp. $NexAbsForms(A)$) to be a function which returns the caller-formulas (resp. abstract-formulas) in $A$. Formally:}
							\ifdefined \NotNeedToReducePages \begin{itemize} \else \begin{itemize}[noitemsep,topsep=0pt] \fi
									\item { $NexCallerForms(A) = \{ X^{c} \phi' \; | \; X^{c} \phi' \in A \}$}
									\item {\changeh{$NexAbsForms(A) = \{ X^{a} \phi' \; | \; X^{a} \phi' \in A \}$}}
								\end{itemize}


								\section{Dynamic Pushdown Networks (DPNs)}
								\subsection{Definitions}
								\label{sec:DPNsDefinition}
								\change{Dynamic Pushdown Networks (DPNs)} is a natural model for multithreaded programs \cite{DBLP:conf/concur/BouajjaniMT05}. To be able to define CARET formulas over DPNs, we must extend \change{this model} to \change{record} whether \change{a transition rule corresponds} to a \textit{call}, \textit{ret} or a \textit{simple} statement (neither call nor ret).
								
								\begin{defn} 
									A Dynamic Pushdown Network (DPN) $\mathcal{M}$ is a set $\{ \mathcal{P}_1, ..., \mathcal{P}_n \}$ s.t. for every $1 \leq i \leq n$,   $ \mathcal{P}_i = (P_i, \Gamma_i, \Delta_i)$ is a Labelled Dynamic Pushdown System (DPDS), where $P_i$ is a finite set of control locations, $P_i \; \cap \; P_j = \emptyset $ for all $ j \neq i$, $\Gamma_i$ is a finite set of \changeh{stack alphabet}, and $\Delta_i$ is  a finite set of transition rules. Rules of $\Delta_i$ are of the following form,  where $p, p_1 \in P_i,  \changenaa{\gamma, \gamma_1, \gamma_2} \in\Gamma_i, \omega_1 \in \Gamma_i^*$, $\changez{d} \in \{\changez{\Box}, p_s \omega_s \; | \; p_s \omega_s \in \bigcup_{1 \leq j \leq n}P_j \times \Gamma_j^* \} $:
									
									\ifdefined \NotNeedToReducePages \begin{itemize} \else \begin{itemize}[noitemsep,topsep=0pt] \fi

											\item { $(r_1)$ $   p \gamma   \xrightarrow{call}_i  p_1 \gamma_1 \gamma_2  \vartriangleright d $ }
											\item { $(r_2)$ $   p \gamma   \xrightarrow{ret}_i  p_1 \epsilon  \vartriangleright d $}
											
											\item { $(r_3)$ $   p \gamma   \xrightarrow{int}_i  p_1 \omega_1  \vartriangleright d $}

										\end{itemize}

										
									\end{defn}
									
									\notesh{while $d= p_s \omega_s$ describes \change{a spawn rule (a new process is spawned)} ===. The reviewers said that $d= p_s \omega_s$ is NOT a rule  . I changed }
									

									Intuitively, there are two kinds of transition rules \change{depending on the nature} of \changez{$d$}.  \changeh{A rule with \changenta{a} suffix of the form $\vartriangleright \Box$ is a \change{nonspawn rule (does not spawn a new process)}, while a rule with \changenta{a} suffix $\vartriangleright p_s \omega_s$ describes \change{a spawn rule (a new process is spawned)}}. A \change{nonspawn} step describes pushdown operations of one single process in the network. Roughly speaking, a $call$ statement is described by a rule in the form $ p \gamma  \xrightarrow{call}_i  p_1 \gamma_1 \gamma_2 \vartriangleright d \in \Delta_i$. This rule usually models a statement of the form $\gamma \xrightarrow{call~~proc} \gamma_2$ where	 $\gamma$ is the control point of the program where the function call is made, $\gamma_1$ is the entry point of the called procedure \changenaa{$proc$},
									and $\gamma_2$ is the return point of the \changenaab{call; $p$ \changenaac{and} $p_1$ can be used to encode various information, such as the return values of functions, shared data between procedures, \changenaac{etc.}} A return statement is modeled by a rule \changenr{$(r_2)$} , while a rule \changenr{$(r_3)$} is used to model a \textit{simple} statement (neither a call nor a return). A \change{spawn} step allows in addition the creation of a new process. For instance,  a rule of the form  $p \gamma \xrightarrow{t}_i p_1 \omega_1 \vartriangleright \change{ p_s \omega_s } \in \Delta_i $ where $t \in \{call, ret, int\}$ describes that a process $\mathcal{P}_i$ at control location $p$ and \changez{having $\gamma$ on top of the stack} can (1) change the control location to $p_1$ and modify the stack by replacing $\gamma$ with $\omega_1$ and also (2) create a new instance of a process $\mathcal{P}_j$ ($1 \leq j \leq n$) starting at $p_s \omega_s$. \change{Note that in this case, if $t$ is call, then $\omega_1$ is $\gamma_1 \gamma_2$, and  if t is \textit{ret}, then $\omega_1$ is $\epsilon$.}

									A DPDS $\mathcal{P}_i$ can be seen as a Pushdown System (PDS) if there are no spawn rules in $\Delta_i$. \change{Generally speaking, a DPN \change{consists of} a set of PDSs}  $\{ \mathcal{P}_1, ..., \mathcal{P}_n \}$  running in parallel where each PDS can dynamically spawn new instances of PDSs in the set  $\{ \mathcal{P}_1, ..., \mathcal{P}_n \}$  during the run. An initial local configuration of a newly created instance $p_s \omega_s$ is called a Dynamically Created Local Initial Configuration (DCLIC). \changeh{For every $i \in \{1 ... n\}$, let $\mathcal{D}_i = \{p_s \omega_s  \in \bigcup_{1 \leq j \leq n} P_j \times \Gamma_j^* \; | \;  p \gamma \xrightarrow{t}_i p_1 \omega_1 \vartriangleright \change{ p_s \omega_s }\in \Delta_i \} $ be the set of DCLICs \changenta{that can be} created by the DPDS $\mathcal{P}_i$.}


									A \textit{local configuration} of an instance of a DPDS $\mathcal{P}_i$ is a tuple $p \omega$ where $p \in P_i$ is the control location, $\omega \in \Gamma_i^*$ is the stack content. A \textit{global configuration} of $\mathcal{M}$ is a multiset over $ \bigcup_{1 \leq i \leq n}P_i \times \Gamma_i^*$, in which $p \omega \in P_i \times \Gamma_i^*$ is a \textit{local configuration} of an instance of  $\mathcal{P}_i$  which is running in parallel in the network $\mathcal{M}$. 
									
									A DPDS $\mathcal{P}_i$ defines a transition relation $\changeh{\xRightarrow{}_{i}}$ as follows: if  $p \gamma \changenrpc{\xrightarrow{t}_i} p_1 \omega_1 \vartriangleright d $ then $p \gamma \omega \changenaa{\xRightarrow{}_{i}} p_1 \omega_1 \omega \vartriangleright D$ for every $\omega \in \Gamma_i^*$ \change{where $D = \emptyset$ if $\changez{d = \Box}$, $D = \{p_s \omega_s\}$ if $d = p_s \omega_s$}.  Let  $\changenaa{\xRightarrow{}^*_{\smash{i}}}$ be the transitive and \change{reflexive} closure of $\changenaa{\xRightarrow{}_{i}}$, then, for every $p \omega \in P_i \times \Gamma_i^*$:

									\ifdefined \NotNeedToReducePages \begin{itemize} \else \begin{itemize}[noitemsep,topsep=0pt] \fi
											\item {$p \omega \changeh{\xRightarrow{}^*_{\smash{i}}} p  \omega \vartriangleright \emptyset$ }
											\item{if $p \omega \changeh{\xRightarrow{}^*_{\smash{i}}} p_1  \omega_1 \vartriangleright D_1$ and $p_1 \omega_1 \changeh{\xRightarrow{}^*_{\smash{i}}} p_2 \omega_2 \vartriangleright D_2$, then, $p \omega \changeh{\xRightarrow{}^*_{\smash{i}}} p_2 \omega_2 \vartriangleright D_1 \cup D_2$ }
										\end{itemize}


											A \textit{local run} of an instance of a DPDS $\mathcal{P}_i$ starting at a local configuration $c_0$ is a sequence  $c_0c_1...$ s.t. for every $x \geq 0$, $c_x \in P_i \times \Gamma_i^*$ is a local configuration of  $\mathcal{P}_i$, $c_x \changeh{\xRightarrow{}_{i}}  c_{x+1} \vartriangleright D$ for some $D$. \changenaa{A \textit{global run}  $\rho$ of $\mathcal{M}$ from a global configuration $\mathcal{G} = \{p_0 \omega_0, ..., p_k \omega_k\}$ is a set of local runs (possibly infinite) where each local run describes the execution of one instance of a certain DPDS $\mathcal{P}_i$.} \changenaa{Initially}, $\rho$ consists of \changenaa{$k$} local runs \changenaa{of $k$ instances starting from $\{p_0 \omega_0, ..., p_k \omega_k\}$}, \changenaab{when a new instance is created, a new local run of this instance is added to $\rho$}. For \changenaab{example,} when a DCLIC $c$ is created by a certain local run of  $\rho$, a new local run that \change{starts} at $c$ is added to  $\rho$. \changenaa{Note that from a global configuration, we can obtain a set of global runs because from a local configuration, we can have different local runs.}

										\ifdefined \NotNeedToReducePages	
										
										\begin{myColorForRemovePages}
											\subsection{\changenta{From Concurrent Programs to DPNs}}
											
											\notesh{Please read this carefully and give comments about this. I'm not sure if we should point out the way we model because it's too specific and there is some conflict. Based on this model, the spawn command will always go with $int$ tags}


											\notesnta{Actually, for concurrent program, like C program, I think that there are no notion about global variable for each pushdown processes. I think the global variables are shared between pushdown processes. WHAT DO YOU THINK ABOUT THIS? }

											\notesnta{One problem here is that if we use only one artificial control location, what we present nexts is week because we use labelling function depending only the control location. One solution is that change it to simple valuation but it will make more complex. WHAT do you think}

											\noindent 
											\textbf{Flow Graph Systems.} \changeh{We suppose that a concurrent program are given by a set of flow graphs where each flow graph represent for a procedure in the program. Let \textbf{Proc} be a set of procedures of the program, including \textbf{Main}. We consider the following kinds of primitive statements: a spawn of a new thread: $spawn(proc)$; a $call$ to a simple procedure: $call(proc)$; a $ret$ statement: $ret$; a $simple$ statement (neither $call$ nor $ret$). Let $Stmt$ be the set of all statements. Each procedure $proc \in$ \textbf{Proc} is represented by a control flow graph $G_{proc} = (N_{proc}, E_{proc}, e_{proc})$ where $N_{proc}$ is a set of program points of the procedure $proc$; $E_{proc} \subseteq N_{proc} \times Stmt \times N_{proc}$ is a set of edges annotated by primitive statements; $e_{proc} \in N_{proc}$ is the entry point of the procedure $proc$. }

											\medskip
											\noindent
											\textbf{Flow Graph Systems to DPNs.} Given a flow graph system, we construct a DPN $\mathcal{M} = \{ \mathcal{P}_1, ..., \mathcal{P}_n \}$  as follows:
											
											\ifdefined \NotNeedToReducePages \begin{itemize} \else \begin{itemize}[noitemsep,topsep=0pt] \fi
													\item {We construct a DPDS $\mathcal{P}_{i}$ for every $i \in Main \cup \{ proc \; | \; spawn(proc) \in Stmt\}$}
													\item {\changeh{For each DPDS $\mathcal{P}_{i} = (P_{i}, \Gamma_{i}, \Delta_{i})$: we have only one artificial control location $\neq$. $\Gamma_{i}$ is a set of all pairs $(n, v)$ where $n$ is a node of the flow graph of the current procedure, $v$ is a tuple represents the value of local variables of the procedure that $n$ belongs to. Thus, the topmost stack alphabet contains the current program point and the current values of local variables of the current procedure, while the remaining stack symbols consists of the return point of its caller and the local values at those corresponding points. $\Delta_{i}$ is a set of transition rules which describe computation steps on the flow graph of the current procedure:}

														\ifdefined \NotNeedToReducePages \begin{itemize} \else \begin{itemize}[noitemsep,topsep=0pt] \fi 
																
																\item { $\neq \llparenthesis n_1, loc \rrparenthesis   \xrightarrow{call}_i \neq \llparenthesis e_{proc}, loc' \rrparenthesis  \llparenthesis n_2, loc \rrparenthesis  $ if there is an edge between program points $n_1$ and $n_2$ denoted by a call to procedure $proc$, where $e_{proc}$ is the entry point of the called procedure $proc$, $loc'$ is the initial values of its local variables.}

																\item {$\neq \llparenthesis n_1, loc \rrparenthesis   \xrightarrow{ret}_i \neq \epsilon $ if there is an edge leaving $n_1$ annotated by a return statement.}
																
																\item { $\neq \llparenthesis n_1, loc \rrparenthesis   \xrightarrow{int}_i \neq \llparenthesis n_2, loc \rrparenthesis  \vartriangleright \neq \llparenthesis e_{proc}, loc' \rrparenthesis $ if there is an edge from $n_1$ to $n_2$ contains a spawn statement to create a new thread starting at the procedure $proc$}
																
																\item { $\neq \llparenthesis n_1, loc \rrparenthesis   \xrightarrow{int}_i \neq' \llparenthesis n_2, loc' \rrparenthesis  $ for every edge which is neither a call nor a return statement }

															\end{itemize}

														}
													\end{itemize}

													\changeh{Note that procedures with return values can be modeled by using additional control locations of DPDS and assigning the return value to these control locations.}

												\end{myColorForRemovePages}
												\fi
												
												\ifdefined \UseFullVersion
												\subsection{Examples}
												
												\notesh{I think we should add examples here because the reviewers of iFM seems can not follow a lot of definition}
												
												\noindent
												\textbf{Example 1.}  \changeh{Let us consider a simple example to illustrate various definitions of DPN. Let $\mathcal{M} = \{ \mathcal{P}_1, \mathcal{P}_2, \mathcal{P}_3 \}$ be a DPN where $\mathcal{P}_1$ is a pushdown process  that consists of two procedure (function) $m$ and $g$. Similarly, $\mathcal{P}_2$ includes two procedures $m'$ and $g'$;  $\mathcal{P}_2$ contains two procedures  $m''$ and $g''$.  Let $P_i = \{p_i\}$ for every $i \in \{1..3\}$, $\Gamma_1 = \{m_0, m_1, m_2, m_3, g_0, g_1, g_2\}$, $\Gamma_2 = \{m'_0, m'_1, m'_2, m'_3, g'_0, g'_1, g'_2\}$,  and $\Gamma_3 = \{m''_0, m''_1, m''_2, \allowbreak  m''_3, g''_0, g''_1, g''_2\}$. The transition rules $\Delta_1$, $\Delta_2$ and $\Delta_3$ are defined as follows:}
												
												\[
												\Delta_1=
												\left\{
												\begin{matrix}
												p_1 m_0  \changenrpc{\xrightarrow{int}_i}  p_1  m_1, &  p_1 m_1  \changenrpc{\xrightarrow{call}_i}  p_1 g_0 m_2, &  p_1 m_2  \changenrpc{\xrightarrow{int}_i}  p_1 m_3 \vartriangleright p_2 m'_1 &  p_1 m_3  \changenrpc{\xrightarrow{int}_i}  p_1 m_3\\
												p_1 g_0  \changenrpc{\xrightarrow{int}_i}  p_1  g_1, &  p_1 g_1  \changenrpc{\xrightarrow{int}_i}  p_1  g_2 \vartriangleright p_3 g''_1, &  p_1 g_2  \changenrpc{\xrightarrow{ret}_i}  p_1  \epsilon\\
												\end{matrix}
												\right\}
												\]
												
												\[
												\Delta_2=
												\left\{
												\begin{matrix}
												p_2 m'_0  \changenrpc{\xrightarrow{int}_i}  p_2  m'_1, &  p_2 m'_1  \changenrpc{\xrightarrow{call}_i}  p_2 g'_0 m'_2, &  p_2 m'_2  \changenrpc{\xrightarrow{int}_i}  p_2 m'_3 \vartriangleright p_1 g_0 &  p_2 m'_3  \changenrpc{\xrightarrow{int}_i}  p_2 m'_3\\
												p_2 g'_0  \changenrpc{\xrightarrow{int}_i}  p_2  g'_1, &  p_2 g'_1  \changenrpc{\xrightarrow{int}_i}  p_2  g'_2, &  p_2 g'_2  \changenrpc{\xrightarrow{ret}_i}  p_2  \epsilon\\
												\end{matrix}
												\right\}
												\]

												\[
												\Delta_3=
												\left\{
												\begin{matrix}
												p_3 m''_0  \changenrpc{\xrightarrow{int}_i}  p_3  m''_1, &  p_3 m''_1  \changenrpc{\xrightarrow{call}_i}  p_3 g''_0 m''_2, &  p_3 m''_2  \changenrpc{\xrightarrow{int}_i}  p_3 m''_3  &  p_3 m''_3  \changenrpc{\xrightarrow{int}_i}  p_3 m''_3\\
												p_3 g''_0  \changenrpc{\xrightarrow{int}_i}  p_3  g''_1, &  p_3 g''_1  \changenrpc{\xrightarrow{int}_i}  p_3  g''_2 , &  p_3 g''_2  \changenrpc{\xrightarrow{ret}_i}  p_3  \epsilon\\
												\end{matrix}
												\right\}
												\]

												In this example:
												
												\ifdefined \NotNeedToReducePages \begin{itemize} \else \begin{itemize}[noitemsep,topsep=0pt] \fi
														\item {\changeh{$\mathcal{M}$ consists of three pushdown processes and each processes can dynamically create new instances of other processes, i.e.  $\mathcal{P}_1$ can create new instances of $\mathcal{P}_2$ starting from $p_2 m'_1$ or create new instances of $\mathcal{P}_3$ starting from $p_3 g''_1$}}
														
														\item {\changeh{$\mathcal{G}_1 = \{p_1 m_0, p_2 m'_0, p_3 m''_0\}$, $\mathcal{G}_2 = \{p_1 g_0, p_2 g'_1\}$ or $\mathcal{G}_3 = \{p_1 m_0, p_1 g_0\}$ are global configurations of $\mathcal{M}$}}
														
														\item {\changeh{$\changeh{\uppi} = p_1 m_0 p_1 m_1 p_1 g_0 ...$ is a local run of pushdown process $\mathcal{P}_1$}}
														\item {\changeh{A global run $\rho$ starting from  $\mathcal{G}_3 = \{p_1 m_0, p_1 g_0\}$ includes several (can be unbounded) instances: an instance $I_1$ of $\mathcal{P}_1$ starting from $p_1 m_0$, an instance $I_2$  of $\mathcal{P}_1$ starting form $p_1 g_0$, an instance $I_3$ of $\mathcal{P}_2$ starting from $p_2 m'_1$ created during the run of $I_1$, an instance $I_4$ of $\mathcal{P}_3$ starting from $p_3 g''_1$ created during the run of $I_1$, ... Similarly, we continue adding instances created by $I_2$, $I_3$, $I_4$, ... Thus, the number of instances of a global run can be unbounded. }}
														
														\item {\changeh{The set of DCLICs that can be created by $\mathcal{P}_1$ is $\mathcal{D}_1 = \{p_2 m'_1, p_3 g''_1\}$, while the set of DCLICs of $\mathcal{P}_1$ that can be created by other pushdown processes in the network is $\mathcal{D'}_1 = \{p_1g_0\}$. Similarly, we have $\mathcal{D}_2 = \{p_1 g_0\}$, $\mathcal{D'}_2 = \{p_2 m'_1\}$ and $\mathcal{D}_3 = \emptyset$, $\mathcal{D'}_3 = \{p_3 g''_1\}$}}
													\end{itemize}

													%

													%


													\fi

													\subsection{Single-indexed CARET for DPNs}
													\label{sec:singleIndexedForDPNs}
													
													\change{Given a DPN $\mathcal{M} = \{ \mathcal{P}_1, ..., \mathcal{P}_n \}$, a single-indexed CARET formula $f$  is a formula in the form $\bigwedge_{i=1}^n f_i$ s.t. for every $1 \leq i \leq n$, $f_i$ is a CARET formula in which the satisfiability of its atomic propositions depends only on the DPDS $\mathcal{P}_i$.}

													
													\medskip
													\noindent
													Given a set of atomic propositions $AP$, let $\lambda: \bigcup_{i = 1}^n P_i \rightarrow 2^{AP}$  be a labeling function that \changez{associates} each control location with a set of atomic propositions. 
													
													Let $\changeh{\uppi} = p_0 \omega_0 p_1 \omega_1 ....$ be a local \changenaa{run} of the DPDS $\mathcal{P}_i$. 	We associate to each \changeh{local} configuration $\changez{p_x \omega_x}$ of $\changeh{\uppi}$ a tag
													$\changed{t_x}$ in $\{ call, int, ret \}$ as follows, \change{where \change{$D = \emptyset $ or $D =  \{ p_s \omega_s \}   $}}:
													
													\ifdefined \NotNeedToReducePages \begin{itemize} \else \begin{itemize}[noitemsep,topsep=0pt] \fi

															\item{\changenaa{If $\changez{p_x \omega_x}$ $\xRightarrow{}_i$ $\changez{p_{x+1} \omega_{x+1}} \change{\vartriangleright D}$ corresponds to a transition rule $p \gamma \changenrpc{\xrightarrow{t}_i} p_1 \omega_1 \vartriangleright d $,  then $\changeh{t_x}=t$.}}
															
%
%
															
														\end{itemize}

														Then, we say that $\changeh{\uppi}$ satisfies $f_i$  iff the $\omega$-word $(\lambda(p_0),t_0) (\lambda(p_1),t_1) ...$  satisfies $f_i$. A local configuration $c$ of $\mathcal{P}_i$ satisfies $f_i$ \changenaa{(denoted $ c \vDash f_i$)} iff there exists a local \changenaa{run} $\changeh{\uppi}$ \changenaa{starting} from $c$ such that $\changeh{\uppi}$ satisfies $f_i$. \changenta{If $D$ is} the set of DCLICs created during the \changenaa{run} $\changeh{\uppi}$, then, we write $c \vDash_D f_i$. A DPN $\mathcal{M}$ satisfies a single-indexed CARET formula $f$ iff there exist a global run $\rho$ s.t. for every $1 \leq i \leq n$, each local run of $\mathcal{P}_i$ in $\rho$ satisfies the formula $f_i$.

														\ifdefined \UseFullVersion
														
														\noindent
														\textbf{\changeh{Example 2. Let $f = f_1 \wedge f_2 \wedge f_3$ is a single-indexed formula which we want to check if the DPN $\mathcal{M} $ presented in Example 1 satisfy or not. Then, we need to check if there exists a global run $\rho$ s.t. every instance of $\mathcal{P}_1$ (resp.  $\mathcal{P}_2$ and  $\mathcal{P}_3$) created during the run of $\rho$ satisfies $f_1$ (resp. $f_2$ and $f_3$).  }  }
														
														\fi


														\section{\changez{Applications}}
														
														\notesh{Maybe it is more intersting to  add the FULL application of Bagle HERE. It will make the Motivation more CLEAR. WHAT DO YOU THINK?}
														
														\changee{We show in this section how \changenta{model-checking} single-indexed CARET for DPNs is necessary for concurrent malware detection.}

															

														\notes{Should we ALL change "The malware" to "The worm"  in this section ??????}
														
														\change{Malware detection is nowadays a big challenge. Several malwares are multithreaded programs that involve  recursive procedures and dynamic thread  creation. Therefore, DPNs can be used to model such programs. We show in what follows how single-indexed CARET  for  DPNs can  describe malicious behaviors of concurrent malwares.}

														\change{More precisely, we show how this logic can specify email worms. To this aim,  let us consider a typical  email worm: the worm  Bagle.} Bagle is a multithreaded email worm. In the main thread, one of the first things \change{the worm} does is to register itself into the \change{registry listing} to be started at the boot time. Then, it does some different actions to \change{hide} itself from users. After \change{this}, the malware \change{creates} one thread (named \textit{Thread2}) that \change{listens} on the port 6777 to receive different commands and also \change{allow} the attacker to upload a new file and execute it. This grants the attacker the ability to update new versions for his malware. In addition, the attacker can send a crafted byte sequence to this port to force the malware to kill itself and delete it from the \change{system}. Thus, the attacker can remove his malware remotely. In the next step, the malware \change{creates} one more thread (named \textit{Thread3}) which contacts a list of websites every 10 minutes to announce the infection of the current machine. \change{The} malware \change{sends} the port it is listening \change{to} as well as the IP of the infected machine to these sites. At some point in the program, the malware \change{continues} to spawn a thread named \textit{Thread4} to search on local drives to look for valid email addresses. In this thread, for each email address found, the malware attaches itself and \change{sends itself} to this email address.

														Thus, \change{you} can see that Bagle is a mutithreaded malware with dynamic thread creation, \change{i.e.,} \changez{the main process can \changed{create} threads} to fulfill various tasks. To model Bagle, DPNs is a good candidate since DPNs allow dynamic thread creation. Let $\mathcal{M} = \{\mathcal{P}_1, \mathcal{P}_2, \mathcal{P}_3, \mathcal{P}_4 \}$ be a model of Bagle \changez{where $\mathcal{P}_1$ is a PDS that represents the main process of the malware}; $\mathcal{P}_2, \mathcal{P}_3, \mathcal{P}_4 $ \changez{are PDSs that model the code segments corresponding to} Thread1, Thread2, Thread3 respectively. Note that $\mathcal{P}_2, \mathcal{P}_3, \mathcal{P}_4 $ are designed to execute specific tasks, while $\mathcal{P}_1$ is \changez{a main process} \changez{able to dynamically create an arbitrary number of instances of $\mathcal{P}_2, \mathcal{P}_3, \mathcal{P}_4 $} \change{to} fulfill tasks in need.

														%

														
														\medskip
														\noindent
														\change{We show now how the malicious behavior of the different  threads can be described by a CARET formula.} Let \change{us} start with the main \changez{process}.  The typical behaviour of this \changez{process} is to add its own executable name to the registry listing so that it can be started at the boot time. To do this, the malware \change{needs} to invoke \change{the API function} $GetModuleFileNameA$ with $0$ and $x$ as parameters. \changez{$GetModuleFileNameA$ will put the file name of its current executable on the memory address pointed by x}. After that, the malware calls \change{the API function} $RegSetValueExA$ with the same $x$ as parameter. $RegSetValueExA$ will use the file name stored at $x$ to add itself into the registry key listing. This malicious behaviour can be specified by CARET as follows:
														
														\noindent
														\resizebox{\hsize}{!}{$\psi_{1} = \boldsymbol{\bigvee_{x \in K}} F^g (call(GetModuleFileNameA) \wedge 0 x \Gamma^* \wedge F^a(call(RegSetValueExA) \wedge x \Gamma^*))$}
														
														\noindent
														where the $\bigvee$ is taken over all possible memory addresses $x$ \changez{over domain $K$}.
														
														
														\done{explain $x \Gamma^*$ here, why use $F^g$, $F^a$}
														
														\change{Note that parameters are passed via the stack in binary programs. For succinctness, we use regular variable expression $x\Gamma^*$ (resp.  $0 x \Gamma^*$) to describe the requirement that $x$ (resp.  $0 x$) is on top of the stack.} Then, this formula states that there is a call to the API \textit{GetModuleFileNameA} with $0$ and $x$ on the top of the stack (i.e., with $0$ and $x$ as parameters), followed by a call to the API $RegSetValueExA$ with $x$ on the top of the stack. Using the operator $F^a$ guarantees that \changed{\textit{RegSetValueExA} is called after \textit{GetModuleFileNameA} terminates.}
														

														\changee{Similarly, the malicious behaviors of the Threads 2, 3 and 4 can be described by CARET formulas $\psi_2$, $\psi_3$ and $\psi_4$ respectively . }
														
%
%
														
														%
														%

														\changea{Thus, the malicious behavior of the concurrent worm Bagle can be described by the single-indexded CARET formula $\psi = \psi_1 \wedge \psi_2 \wedge \psi_3 \wedge \psi_4$.}

														\section{Single-indexed CARET model-checking for DPNs}
														
														\done{Rewrite this}
														
														In this section, we consider the CARET model-checking problem of DPNs. Let $\lambda
														: \bigcup_{i = 1}^n P_i \rightarrow 2^{AP}$ be a labeling function that associates each control location with a set of atomic propositions. Let $\mathcal{M} = \{ \mathcal{P}_1, ..., \mathcal{P}_n \}$ be a DPN, $f = \bigwedge_{i=1}^n f_i$ be a single-indexed CARET formula.

														\subsection{B\"{u}chi DPNs (BDPNs)}


														\begin{defn}
															A B\"{u}chi DPDS (BDPDS) is a tuple $\mathcal{BP}_i = (P_i, \Gamma_i, \Delta_i, F_i)$ s.t. $\change{\mathcal{P}_i} = (P_i, \Gamma_i, \Delta_i)$ is a DPDS, $F_i \subseteq P_i$ is the set of accepting control locations. \change{A run of a BDPDS is accepted iff it visits infinitely often some control locations in $F_i$.}
														\end{defn}

														\begin{defn}
															A \textit{Generalized B\"{u}chi DPDS} (GBDPDS) is a tuple $\mathcal{BP}_i = (P_i, \Gamma_i, \Delta_i, F_i)$, where $\change{\mathcal{P}_i} = (P_i, \Gamma_i, \Delta_i)$ is a DPDS and $F_i = \{ F_1, ..., F_k \}$ is a set of sets of accepting control locations. A run of a GBDPDS is accepted iff it visits infinitely often some control locations in $F_j$ for every $1 \leq j \leq k$.
														\end{defn}

														
														\change{Given a BDPDS or a GBDPDS $\mathcal{BP}_i  = (P_i, \Gamma_i, \Delta_i, F_i) $, let $c \in P_i \times \Gamma_i^*$ \changez{be a} local configuration of  $\mathcal{BP}_i$. Then, let  $\mathcal{L}(\mathcal{BP}_i)$ be the set of all pairs $(c, D) \in P_i \times \Gamma^*_i \times 2^{\mathcal{D}_i}$  s.t. $\mathcal{BP}_i$ has an accepting run from $c$ and $D$ is the set of DCLICs generated} during that run. We get the following properties:

														\begin{prop}
															Given a GBDPDS $\mathcal{BP}_i$,
															we can effectively compute a BDPDS $\mathcal{BP'}_i$ s.t. $\mathcal{L} (\mathcal{BP}_i) = \mathcal{L} (\mathcal{BP'}_i)$.
														\end{prop}
														
														\changenaac{This result comes from the fact that we can translate a GBDPDS to a corresponding BDPDS by applying the similar approach as the translation from a Generalized B\"{u}chi automaton to a corresponding B\"{u}chi automaton \changenaac{\cite{clark99}}.}

														
														%
														%

														\begin{defn}
															A B\"{u}chi Dynamic Pushdown Network (BDPN) is a set $\{ \mathcal{BP}_1, ..., \mathcal{BP}_n\}$ s.t. for every $1 \leq i \leq n$,   $ \mathcal{BP}_i = (P_i, \Gamma_i, \Delta_i, F_i)$ is a  BDPDS.  \change{A (global) run $\rho$ of a BDPN is accepted iff all local runs in $\rho$ are accepting (local) runs.}

														\end{defn}

														\begin{defn}
															A Generalized B\"{u}chi Dynamic Pushdown Network (GBDPN) is a set $\{ \mathcal{BP}_1, ..., \mathcal{BP}_n\}$ s.t. for every $1 \leq i \leq n$,   $ \mathcal{BP}_i = (P_i, \Gamma_i, \Delta_i, F_i)$ is a  GBDPDS. \change{A (global) run $\rho$ of a GBDPN is accepted iff all local runs in $\rho$ are accepting (local) runs.}

														\end{defn}

														
															\change{Given a BDPN or a GBDPN \changez{$\mathcal{BM} = \{ \mathcal{BP}_1, ..., \mathcal{BP}_n\} $, let  $\mathcal{L}(\mathcal{BM})$} be the set of all global configurations $\mathcal{G}$ s.t. $\mathcal{BM}$ has an accepting run from $\mathcal{G}$. We get the following properties:}

														\begin{prop}
															\label{prop:GBDPN2BDPN}
															Given a GBDPN $\mathcal{BM}$,
															we can effectively compute a \changed{BDPN} $\mathcal{BM}'$ s.t. $\mathcal{L} (\mathcal{BM}) = \mathcal{L} (\mathcal{BM}')$.
														\end{prop}

														\changez{This result is obtained due to the fact that we can translate each GBDPDS in $\mathcal{BM}$ to a corresponding BDPDS in  $\mathcal{BM}'$.}

														\medskip
														\noindent
														Given a BDPN  $\mathcal{BM} = \{ \mathcal{BP}_1, ..., \mathcal{BP}_n\}$ where $\mathcal{BP}_i = (P_i, \Gamma_i, \Delta_i, F_i)$. Let $I(c)$ be the index $i$ of the local configuration $c \in P_i \times \Gamma_i^*$. Let  $\mathcal{D} = \bigcup_{i=1}^n \mathcal{D}_i$. Then, we get the following theorem:

														\begin{theorem}
															\label{theorem:memberShipOfBDPN}
															\changenrp{\cite{DBLP:conf/aplas/SongT13,DBLP:journals/corr/SongT16}} The membership problem of a BDPN is decidable in time $O(\Sigma_{i=1}^n | \Delta_i | . \allowbreak | \Gamma_i| . |P_i|^3 . 2^{|\mathcal{D}_i|} + \Sigma_{c \in \mathcal{D}} (|c| . |P_{I(c)}|^3 . |\Gamma_{I(c)} | .  2^{2 |\mathcal{D}_{I(c)}|} + |\mathcal{D}|^2 . 2^{|\mathcal{D}|})$.
														\end{theorem}

														\changez{Thus, from Proposition \ref{prop:GBDPN2BDPN} and Theorem \ref{theorem:memberShipOfBDPN}, we get that the membership problem of a GBDPN is decidable.}
														
														
														\begin{theorem}
															The membership problem of \changec{GBDPNs} is decidable.
														\end{theorem}
														\notes{I add Theorem here}


														\done{Explain symbols here}

														
														\subsection{\changez{From CARET model checking of DPNs to the membership problem in BDPNs}}
														\label{sec:DPN_GBDPDSComputation}
														
														Given a local run $\changend{\uppi}$, let \changez{$\vartheta(\changend{\uppi})$} be the index of the DPDS corresponding to $\changend{\uppi}$. Let $\mathcal{G}$ be an initial global configuration of \changenaa{the DPN} $\mathcal{M}$, then we say that $\mathcal{G}$ satisfies $f$ iff $\mathcal{M}$ has a global run $\rho$ starting from $\mathcal{G}$  s.t. every local run $\changend{\uppi}$ in $\rho$ satisfies $f_{\vartheta(\changend{\uppi})}$. Determining whether $\mathcal{G}$ satisfies $f$ is a non-trivial problem since the number of global runs can be unbounded and the number of local runs of each global run can also be unbounded. Note that it is not sufficient to check whether every pushdown process $\mathcal{P}_i$ satisfies the corresponding CARET formula $f_i$. Indeed, we need to ensure that all instances of $\mathcal{P}_i$ created during a global run \changez{satisfy} the formula $f_i$. Also, it is not correct to check whether all possible instances of $\mathcal{P}_i$ \changenb{satisfy} the formula $f_i$. Indeed, an instance of $\mathcal{P}_i$ should not be checked if it is not created during a global run. \changeh{To solve these problems, we reduce the CARET model-checking problem for DPNs to the membership problem for GBDPNs. To do this, we compute a GBDPN $\mathcal{BM} = \{ \mathcal{BP}_1, ..., \mathcal{BP}_n\} $ where $\mathcal{BP}_i$ ($i \in \{1..n\}$) is a GBDPDS  s.t. (1) the problem of checking whether each instance of $\mathcal{P}_i$ satisfies a CARET formula $f_i$ can be reduced to the \changenaab{membership problem of $\mathcal{BP}_i$}; (2) if $\mathcal{P}_i$ creates a new instance of $\mathcal{P}_j$ starting from $p_s \omega_s$, which requires that $p_s \omega_s \vDash f_j$; $\mathcal{BP}_i$ must also create an instance of $\mathcal{BP}_j$ \changenb{starting from a certain configuration \changenta{(computed from $p_s \omega_s$)} from which $\mathcal{BP}_j$ has an accepting run.} In what follows, we present how to compute such GBDPDSs.}

														

														
														\changenta{Let $Label = \{exit, unexit \}$ (we explain later the need to these labels)}. 	\changeh{Given a DPDS $\mathcal{P}_i$ ($i \in \{1..n\}$), a corresponding CARET formula $f_i$, we define $Initial_i$ as the set of atoms A ($A \in Atoms(f_i)$) such that $f_i \in A$ and $NextCallerFormulas(A) = \emptyset$. \changenta{Our goal is that for every $\mathcal{P}_i$ ($i \in \{1..n\}$), we compute a GBDPDS $\mathcal{BP}_i$ s.t. for every $p \omega \in P_i \times \Gamma_i^*$, $p \omega$ satisfies $f_i$ iff there exists an atom  $A$ where $A \in Initial_i$  s.t. $\mathcal{BP}_i$ has an accepting run from $ \llparenthesis p, A \changeh{,unexit} \rrparenthesis  \omega $}}.

														\medskip
														\noindent
														{\bf GBDPDSs Computation.}

														\changeh{Let us fix a DPDS \changeh{$\mathcal{P}_i = (P, \Gamma, \Delta)$} in the DPN $\mathcal{M}$, a CARET formula $f_i$ in   $f = \bigwedge_{i=1}^n f_i$ corresponding to the DPDS $\mathcal{P}_i$. In this section, we show how to compute such a \changez{GBDPDS}  $\mathcal{BP}_i$ corresponding to $\mathcal{P}_i$. \changenta{Given a local configuration $p \omega$, let $\delta(p \omega)$ be the index of the DPDS corresponding to $p \omega$.} We define $\mathcal{BP}_i = (P', \Gamma', \Delta', F)$ as follows:}


														\ifdefined \NotNeedToReducePages \begin{itemize} \else \begin{itemize}[noitemsep,topsep=0pt] \fi
																\item{\changeh{$P' = \{ \llparenthesis p, A, l \rrparenthesis \; | \; p \in P, l \in \changeh{Label}, A \in Atoms(f_i)$} and $A \cap AP = \lambda (p) $ \} is the finite set of control locations of \changeh{$\mathcal{BP}_{i}$}}
																\item{$\Gamma' = \Gamma \cup (\Gamma \times Atoms(f_i) \changeh{\times \changeh{Label}})$ is the finite set of stack symbols of \changeh{$\mathcal{BP}_{i}$}.}
																
																%
															\end{itemize}
															\medskip
															The transition relation $\Delta'$ of \changeh{$\mathcal{BP}_{i}$} is the smallest set of transition rules satisfying the following:
															\ifdefined \NotNeedToReducePages \begin{itemize} \else \begin{itemize}[noitemsep,topsep=0pt] \fi
																	\item{$(\alpha_{1})$ for every $\changene{ p \gamma } \changenrpc{\xrightarrow{call}_i} \changene{ q \gamma' \gamma'' } \changeh{\vartriangleright d} \in \Delta$: $\changenrlangle{\changeh{\llparenthesis p, A, l \rrparenthesis} \gamma } \changenrpd{\longrightarrow_i}\changenrlangle{\changeh{\llparenthesis q, A', l' \rrparenthesis} \gamma' \changeh{\llparenthesis \gamma'', A, l \rrparenthesis} } \changeh{\vartriangleright d_0} \in \Delta'$} for every $A, A' \in Atoms(f_i)$\changeh{; $l,l' \in \changeh{Label}$} such that:
																	\ifdefined \NotNeedToReducePages \begin{itemize} \else \begin{itemize}[noitemsep,topsep=0pt] \fi
																			\item{$(\beta_{0})$ $A \cap \{call, ret, int \} = \{ call \} $}
																			\item{$(\beta_{1})$ $A \cap AP = \lambda (p) $}
																			\item{$(\beta_{2})$ $ A' \cap AP = \lambda (q) $}
																			\item{$(\beta_{3})$ $GlNext(A, A')$}
																			
																			\item{$(\beta_{4})$ $CallerNext(A', A) $}
																			\item{\changeh{$(\beta_{5})$  $l' = \changeh{unexit}$ implies ($l = \changeh{unexit}$ and $NexAbsForms(A) = \emptyset$)} }

																			\item {\changeh{$(\beta_{6})$ $d_0 = \Box$ if $d = \Box$; $d_0 = \llparenthesis p_s, A_0, unexit \rrparenthesis \omega_s$ where $A_0 \in Initial_{\delta(p_s \omega_s)}$ if $d = p_s \omega_s$ }  }
																			

																		\end{itemize}
																		\item{ $(\alpha_{2})$ for every $\changene{ p \gamma } \changenrpc{\xrightarrow{ret}_i} \changene{ q \epsilon } \changeh{\vartriangleright d} \in \Delta$: }
																		\ifdefined \NotNeedToReducePages \begin{itemize} \else \begin{itemize}[noitemsep,topsep=0pt] \fi
																				\item{$(\alpha_{2.1})$ $\changenrlangle{\changeh{\llparenthesis p, A, \changeh{exit} \rrparenthesis} \gamma} \changenrpd{\longrightarrow_i}\changenrlangle{\changeh{\llparenthesis q, A', l' \rrparenthesis} \epsilon } \changeh{\vartriangleright d_0} \in \Delta'$ for every $A, A' \in Atoms(f_i)$ \changeh{$;l,l' \in \changeh{Label}$} such that:}
																				\ifdefined \NotNeedToReducePages \begin{itemize} \else \begin{itemize}[noitemsep,topsep=0pt] \fi
																						\item{$(\beta_{0})$ $A \cap \{call, ret, int \} = \{ ret \} $}
																						\item{$(\beta_{1})$ $A \cap AP = \lambda (p) $}
																						\item{$(\beta_{2})$ $A' \cap AP = \lambda (q) $}
																						\item{$(\beta_{3})$ $GlNext(A, A')$}
																						\item{\changeh{$(\beta_{4})$ $NexAbsForms(A) = \emptyset$}}
																						\item {\changeh{$(\beta_{5})$ $d_0 = \Box$ if $d = \Box$; $d_0 = \llparenthesis p_s, A_0, unexit \rrparenthesis \omega_s$ where $A_0 \in Initial_{\delta(p_s \omega_s)}$ if $d = p_s \omega_s$ }  }

																					\end{itemize}
																					\item{$(\alpha_{2.2})$ $\changenrlangle{\changeh{\llparenthesis q, A', l' \rrparenthesis, \llparenthesis \gamma_0, A_0, l_0 \rrparenthesis } \rangle \changenrpd{\longrightarrow_i}\langle \changeh{\llparenthesis q, A', l' \rrparenthesis} \gamma_0 } \in \Delta'$ for every $\gamma_0 \in \Gamma, A_0, A' \in Atoms(f_i)$\changeh{; $l', l_0 \in \changeh{Label}$} such that:}
																					\ifdefined \NotNeedToReducePages \begin{itemize} \else \begin{itemize}[noitemsep,topsep=0pt] \fi
																							\item{$(\beta_{6})$ $AbsNext(A_0, A')$}
																							\item{$(\beta_{7})$ $NexCallerForms(A') = NexCallerForms(A_0)$}
																							\item{$(\beta_{8})$ $A' \cap AP = \lambda (q) $}
																							\item{\changeh{$(\beta_{9})$ $l_0 = l'$}}

																						\end{itemize}
																					\end{itemize}
																					\item{ $(\alpha_{3})$ for every $\changene{ p \gamma } \changenrpc{\xrightarrow{int}_i} \changenrlangle{q \change{\omega} } \changeh{\vartriangleright d} \in \Delta$: $\changenrlangle{\changeh{\llparenthesis p, A, l \rrparenthesis} \gamma} \changenrpd{\longrightarrow_i}\changenrlangle{\changeh{\llparenthesis q, A', l \rrparenthesis} \change{\omega} } \changeh{\vartriangleright d_0} \in \Delta'$ for every $A, A' \in Atoms(f_i)$, \changeh{$l \in \changeh{Label}$} such that:}
																					\ifdefined \NotNeedToReducePages \begin{itemize} \else \begin{itemize}[noitemsep,topsep=0pt] \fi
																							\item{$(\beta_{0})$ $A \cap \{call, ret, int \} = \{ int \} $}
																							\item{$(\beta_{1})$ $A \cap AP = \lambda (p) $}
																							\item{$(\beta_{2})$ $A' \cap AP = \lambda (q) $}
																							\item{$(\beta_{3})$ $GlNext(A, A')$}
																							\item{$(\beta_{4})$ $AbsNext(A, A')$}
																							\item{$(\beta_{5})$ $NexCallerForms(A) = NexCallerForms(A') $}
																							\item {\changeh{$(\beta_{6})$ $d_0 = \Box$ if $d = \Box$; $d_0 = \llparenthesis p_s, A_0, unexit \rrparenthesis \omega_s$ where $A_0 \in Initial_{\delta(p_s \omega_s)}$ if $d = p_s \omega_s$ }  }
																						\end{itemize}
																						
%

																					\end{itemize}

																					\changeh{Let \changex{$cl_{U^g} ( f_i ) = \{ \phi_1 U^g \chi_1, ..., \phi_k U^g \chi_k \}$} and $cl_{U^a} ( f_i ) = \{ \xi_1 U^a \tau_1, ..., \xi_{k'} U^a \tau_{k'} \}$ be the set of $U^g$-formulas and $U^a$-formulas of $Cl(f_i)$ respectively. The generalized B\"{u}chi accepting condition $F$ of $\mathcal{BP}_{i}$ is defined as: $F = \{F_1\} \cup F_2 \cup F_3 $ where }
																					
																					\ifdefined \NotNeedToReducePages \begin{itemize} \else \begin{itemize}[noitemsep,topsep=0pt] \fi
																							 
																							\item {\changeh{$F_1 = P \times \changenrpd{Atoms(f_i)} \changeh{\times \{unexit\}}$}}

						\item {\changeh{$F_2 = \{ F^g_1, ..., F^g_{k} \}$ where  $F^g_x = \{P \times F_{\phi_x U^g \chi_x} \times \changeh{Label}\} $ where $F_{\phi_x U^g \chi_x} = \{ A \in Atoms(f_i) \; | \;$ if $\phi_x U^g \chi_x \in A$ then $\chi_x \in A \}$   for every $1 \leq x \leq k$.}}
						
						\item {\changeh{$F_3 = \{ F^a_1, ..., F^a_{k'} \}$ where  $F^a_x = \{P \times F_{\xi_x U^a \tau_x} \times \changeh{\{unexit\}}\}$ where $F_{\xi_x U^a \tau_x} = \{ A \in Atoms(f_i) \; | \;$ if $\xi_x U^a \tau_x \in A$ then $\tau_x \in A \}$  for every $1 \leq x \leq k'$.}}
					\end{itemize}

																						\notesnta{can you please check  this intuition carefully?}

																						\medskip
																						\noindent
																						\ifdefined \NotNeedToReducePages	

																						\else
																						\noindent
																						\changenrp{Given a configuration $p \omega$, let $\mathscr{P}(p \omega)$ be the procedure to which $p \omega$ belongs. For example, in Figure \ref{fig:nextAbstract}, $\mathscr{P}(p_{x+1} \omega_{x+1}) = proc$, ..., $\mathscr{P}(p_{y-1} \omega_{y-1}) = proc$.}
																						\fi
																						\changez{Intuitively, we compute $\mathcal{BP}_i $ as a kind of product of $\mathcal{P}_i$ and $f_i$ which ensures that: for every $p \omega \in P_i \times \Gamma_i^*$, $p \omega$ satisfies $f_i$ \changenta{iff there exists an atom \changenrpb{ $A \in Intial_i$ s.t. $\mathcal{BP}_i$ has an accepting run from $ \llparenthesis p, A \changeh{,unexit} \rrparenthesis  \omega $. To do this, we encode atoms of $f_i$ into control locations of $\mathcal{P}_i$.}} The form of control locations of $\mathcal{BP}_i$ is $\llparenthesis p, A \changeh{,l} \rrparenthesis$ where $A$ contains all sub formulas of $f_i$ which are satisfied at the configuration $ p \omega $\changeh{, $l$ is a label to \changenrpc{determine whether the execution of the procedure of $\changenrlangle{p \omega }$, $\mathscr{P} (\changenrlangle{p \omega })$ \ifdefined \NotNeedToReducePages (as defined in Section \ref{sec:CARETDefineAtomDefine})\else\fi}, \changenta{terminates in the path} $\changend{\uppi}$. A configuration $\changenrlangle{p \omega }$ \changenta{labeled with} $\changeh{exit}$ means that the execution of $\mathscr{P} (\changenrlangle{p \omega })$ is finished \changeh{in} $\changend{\uppi}$, i.e., the run $\changend{\uppi}$ will run through the procedure $\mathscr{P} (\changenrlangle{p \omega })$, \changeh{reaches} its ret statement and \changeh{exits} $\mathscr{P} (\changenrlangle{p \omega })$ after that. On the contrary, $\changenrlangle{p \omega }$ \changenta{labeled with} $\changeh{unexit}$ means that \changeh{in $\changend{\uppi}$, the execution of the procedure $\mathscr{P} (\changenrlangle{p \omega })$ never terminates}, i.e., the run $\changend{\uppi}$ will be \changeh{stuck in and never exits} the procedure $\mathscr{P} (\changenrlangle{p \omega })$}}. Let \change{$\changend{\uppi} =  p_0 \omega_0 p_1  \omega_1  ... $ be a run of $\mathcal{P}_i$} and 
																						\change{$  \llparenthesis p_0, A_0 \changeh{,l_0} \rrparenthesis \omega_0 \llparenthesis p_1, A_1 \changeh{,l_1} \rrparenthesis \omega_1  .... $}be a
																						corresponding run of $\mathcal{BP}_i$. \changenta{We give in what follows the intuition behind our construction.}

																						\notesnta{I used $d_0$ for all Rules, do you think that I should you different like $d_0$, $d'_0$, $d''_0$ in different rules}

																						\ifdefined \NotNeedToReducePages	
																						
																						\begin{myColorForRemovePages}
																							
																							\medskip
																							\noindent
																							\textbf{Encoding atoms to control locations.} Firstly, we need to ensure that $\mathcal{BP}_i$ has an accepting \changec{(local)} run
																							from $\changed{ \llparenthesis p_x, A_x \rrparenthesis\omega_x }$ \changec{iff $\changec{p_x \omega_x}$ satisfies $\phi$  (denoted $\changec{p_x \omega_x} \vDash \phi$)} for every $\phi \in A_x$. To ensure this, in rules $(\alpha_1)$, $(\alpha_2)$ and $(\alpha_3)$, the first class of conditions $(\beta_0)$ ensures that the tags
																							$\{ call, ret, int \}$ assigned to each configuration \change{of the run} are guessed correctly. The second class of conditions $(\beta_1)$ and $(\beta_2)$ \change{expresses} that for
																							every $e \in AP$, $(\changend{\uppi}, x) \vDash e$ iff $e \in \lambda(p_x)$, and the class of conditions $(\beta_3)$ \change{expresses} that $(\changend{\uppi}, x) \vDash X^g \phi'$ iff
																							$(\changend{\uppi}, x+1) \vDash \phi'$. Now, let us consider the two most delicate cases:

																							\begin{figure*}
																								\centering
																								\begin{tikzpicture}[xscale=1.3, yscale=1.3]
																								\tikzset{lineStyle/.style={blue, ultra thick}};
																								\tikzset{myDot/.style={blue, fill = yellow, thick}};
																								\tikzset{myRectangleNode/.style={rectangle, thick, draw= black, below right, black}};
																								\draw [->, lineStyle, dashed] (0, 0) -- (1,0);
																								\draw [->, lineStyle] (1,0) -- (2,0) -- (2.5, -1) -- (3.5, -1) -- (4.5, -1);
																								\draw [->, lineStyle, dashed] (4.5, -1) -- (5.5, -1);
																								\draw [->, lineStyle] (5.5, -1) -- (6.5, -1) -- (7, 0) -- (8, 0);
																								\draw [->, lineStyle, dashed] (8, 0) -- (9, 0);
																								\draw [myDot] (0,0) circle [radius=0.04];
																								\draw [myDot] (1,0) circle [radius=0.04];
																								\draw [myDot] (2,0) circle [radius=0.04];
																								\draw [myDot] (2.5, -1) circle [radius=0.04];
																								\draw [myDot] (3.5, -1) circle [radius=0.04];
																								\draw [myDot] (4.5, -1) circle [radius=0.04];
																								\draw [myDot] (5.5, -1) circle [radius=0.04];
																								\draw [myDot] (6.5, -1) circle [radius=0.04];
																								\draw [myDot] (7,0) circle [radius=0.04];
																								\draw [myDot] (8,0) circle [radius=0.04];
																								\draw [myDot] (9,0) circle [radius=0.04];
																								\node[below left](call) at (2,0) {$call$};
																								
																								\node[above] at (0,0) {$A_{0}$};
																								\node[above right](AtomA_call) at (2,0) {$A_{x}$};
																								\node[above right]() at (2.5,-1) {$A_{x+1}$};
																								
																								\node[above left](proc-entry) at (2.3,-1) {\changeh{$proc$}};
																								\node[above left](ret) at (6.5,-1) {$ret$};
																								\node[below right](return-point) at (7,0) {\textit{return-point}};
																								\node[above left](AtomA0_pass) at (7,0) {$\llparenthesis \gamma'', A_{x} \rrparenthesis$};
																								\draw[dotted, red, thick, ->] (AtomA_call) .. controls(4,0.5) .. (AtomA0_pass);
																								\node[below right](encoded) at (3,0.3) {encoded $\&$ passed down};
																								\node[below]() at (0,0) {$p_0 \omega_0 $};
																								\node[below right]() at (2,0) {$\changec{p_x \omega_x}$};
																								\node[below]() at (2.5,-1) {$\changec{p_{x+1} \omega_{x+1}}$};
																								\node[below]() at (6.5,-1) {$\changec{p_{y-1} \omega_{y-1}}$};
																								\node[above right]() at (7,0) {$p_y \omega_y$};
																								
																								\node[ below left]() at (6.9,0) {$A_{y}$};

																								\end{tikzpicture}
																								\caption{Case of $X^a \phi' \in A_x$ }
																								\label{fig:nextAbstract}
																							\end{figure*}

																							\begin{enumerate}
																								\item
																								If $\phi = X^{a} \phi' \in A_x$. There are two possibilities:
																								
																								\ifdefined \NotNeedToReducePages \begin{itemize} \else \begin{itemize}[noitemsep,topsep=0pt] \fi
																										
																										\item
																										{ $\changec{p_x \omega_x}$ $\changec{\changend{\xRightarrow{}_{i}}}$ $\changec{p_{x+1} \omega_{x+1}} \changenta{\vartriangleright d_0}$ corresponds to a call statement.
																											Let us consider Figure \ref{fig:nextAbstract} to explain this case. Let $\changec{ p_y\omega_y }$ be the abstract-successor of $\changec{p_x \omega_x}$.
																											$(\changend{\uppi}, x) \vDash X^{a} \phi' $ iff $(\changend{\uppi}, y) \vDash \phi'$. Thus, we must have $\phi' \in A_{y}$.
																											This is ensured by rules $\alpha_1$ and $\alpha_2$: rules $\alpha_1$ allow to record $X^a \phi'$ in the return point of the call, and rules $\alpha_2$ allow
																											to extract and validate $\phi'$ when the return-point is reached. In what follows, we show in more details how this works:
																											Let $\changec{p_x \gamma} \xrightarrow{call}_i \changec{p_{x+1}  \gamma' \gamma'' \vartriangleright d}$ be the rule associated with the transition $\changec{p_x \omega_x}$ $\changec{\changeh{\xRightarrow{}_{i}}}$ $\changec{p_{x+1} \omega_{x+1}} \changenta{\vartriangleright d_0}$, then we have $\omega_x = \gamma \omega'$ and $\omega_{x+1} = \gamma' \gamma'' \omega'$. Let $\changec{p_{y-1} \omega_{y-1}}$ $\changec{\changeh{\xRightarrow{}_{i}}}$ $\changec{p_{y} \omega_{y}} \changenta{\vartriangleright d_0}$ be the transition that corresponds to the $ret$ statement of this call.
																											Let then $\changec{p_{y-1} \beta} \xrightarrow{ret}_i \changec{p_y \epsilon} \vartriangleright d \in \Delta$ be the corresponding return rule.
																											Then, we have necessarily $\omega_{y-1} = \beta \gamma'' \omega'$, since as explained in Section \ref{sec:DPNsDefinition},
																											$\gamma''$ is the return address of the call. After applying this rule, $\omega_{y} = \gamma'' \omega'$.
																											In other words, $\gamma''$ will be the topmost stack symbol at the corresponding return point of the call.
																											So, in order to recover $\phi'$ in $A_{y}$, we proceed as follows:
																											At the call $\changec{p_x \gamma} \xrightarrow{call}_i \changec{p_{x+1}  \gamma' \gamma'' \vartriangleright d}$, we encode $A_x$ into $\gamma''$ by the rule
																											$(\alpha_1)$ stating that $\changec{\llparenthesis p_x, A_x \changeh{, l} \rrparenthesis \gamma} \changenrpd{\longrightarrow_i} \llparenthesis p_{x+1}, A_{x+1} \changeh{, l'}  \rrparenthesis \gamma' 
																											\llparenthesis \gamma'', A_x \changeh{, l} \rrparenthesis  \changenb{\vartriangleright d_0} \in \Delta'$. 
																											This allows to record $X^a \phi'$ in the corresponding return point of the stack.
																											After that, $\llparenthesis \gamma'', A_x \changeh{,l} \rrparenthesis $ will be the topmost stack symbol at the corresponding return-point of this call. 
																											At the return-point, the condition \changencd{$(\beta_6)$} in $(\alpha_{2.2})$ stating that $AbsNext(A_x, A_y)$ and the fact that $\phi = X^{a} \phi' \in A_x$
																											imply that $\phi' \in A_y$. 
																											
																											\item{ $\changec{p_x \omega_x}$ $\changec{\changend{\xRightarrow{}_{i}}}$ $\changec{p_{x+1} \omega_{x+1}} \changenta{\vartriangleright d_0}$ corresponds to a simple statement. Then, the abstract successor of $\changec{p_x \omega_x}$ is $\changec{p_{x+1} \omega_{x+1}}$ (see Figure \ref{fig:nextCaller2}).
																												$(\changend{\uppi}, x) \vDash X^{a} \phi' $ iff $(\changend{\uppi}, x+1) \vDash \phi'$. Thus, we must have $\phi' \in A_{x+1}$. This is ensured by condition $(\beta_4)$ in $(\alpha_3)$ stating that $AbsNext(A_x, A_{x+1}) = true$
																											}
																											
																										}
																									\end{itemize}
																									
																									\item
																									The other delicate case is when $\phi = X^{c} \phi' \in A_x$. This means that $(\changend{\uppi}, x) \vDash X^{c} \phi'$. This case is handled by the conditions $(\beta_4)$ in $(\alpha_1)$, \changencd{$(\beta_7)$} in $(\alpha_2)$ and $(\beta_5)$ in $(\alpha_3)$. Let us consider the example in Figure \ref{fig:nextCaller2} to illustrate this case. In this figure, the caller-successor of $\changec{p_x \omega_x}$ is $\changec{p_{x-t} \omega_{x-t}}$. Thus, $(\changend{\uppi}, x) \vDash X^{c} \phi' $ iff
																									$(\changend{\uppi}, x-t) \vDash \phi'$. Then we need to ensure \change{that} $\phi' \in A_{x-t}$. This is done as follows:
																									\ifdefined \NotNeedToReducePages \begin{itemize} \else \begin{itemize}[noitemsep,topsep=0pt] \fi
																											\item{$\changec{p_{x-1} \omega_{x-1}}$ $\changec{\changeh{\xRightarrow{}_{i}}}$ $\changec{p_{x} \omega_{x}} \changenta{\vartriangleright d_0}$ corresponds to a simple statement, so we require $NexCallerForms(A_{x-1}) = NexCallerForms(A_{x})$ (by the condition $(\beta_5)$ in $(\alpha_{3})$). This implies $X^{c} \phi' \in A_{x-1}$. Similarly, we have $X^c \phi' \in A_{x-y_2}$.}
																											\item{$\changec{ p_{x-y_1} \omega_{x-y_1} }$ and $\changec{ p_{x-y_2} \omega_{x-y_2} }$ is a pair of call and return-point. Then, by applying
																												the condition \changencd{$(\beta_7)$} in $(\alpha_{2.2})$, we have $NexCallerForms(A_{x-y_1}) = NexCallerForms(A_{x-y_2})$. This implies $X^{c} \phi' \in A_{x-y_1}$.}
																											
																											\item{The transitions from $\changec{p_{x-(t-1)} \omega_{x-(t-1)}}$ to \change{$\changec{p_{x- y_1} \omega_{x-y_1}}$} correspond to simple statements. 
																												By the condition $(\beta_5)$ in $(\alpha_{3})$, we obtain $X^c \phi' \in A_{x-(t-1)}$}
																											
																											\item{$\changec{p_{x-t} \omega_{x-t}}$ $\changec{\changeh{\xRightarrow{}_{i}}}$ $\changec{p_{x-(t-1)} \omega_{x-(t-1)}} \changenta{\vartriangleright d_0}$ corresponds to a call statement, so we require \change{$CallerNext(A_{x-(t-1)}, A_{x-t})$} (by the condition $(\beta_4)$ in $(\alpha_{1})$) which means that if $X^{c} \phi' \in A_{x-(t-1)}$ then $\phi' \in A_{x-t}$. }
																										\end{itemize}
																										
																									\end{enumerate}

																									\begin{figure*}
																										\centering
																										
																										\begin{tikzpicture}[xscale=1.2, yscale=1.3]
																										\tikzset{lineStyle/.style={blue, ultra thick}};
																										\tikzset{myDot/.style={blue, fill = yellow, thick}};
																										\tikzset{myRectangleNode/.style={rectangle, thick, draw= black, below right, black}};
																										
																										\draw [lineStyle, dashed, ->] (0, 0) -- (1,0);
																										
																										\draw [lineStyle] (1,0) -- (2,0);
																										\draw [lineStyle, ->] (2,0) -- (2.5, -1);
																										
																										\draw [lineStyle, dashed] (2.5, -1) -- (3.5, -1);
																										\draw [lineStyle, dashed] (3.5, -1) -- (4.5, -1);

																										\draw [lineStyle, ->] (4.5, -1) -- (5, -2);
																										\draw [lineStyle, dashed, ->] (5, -2) -- (6, -2);
																										\draw [lineStyle, ->] (6, -2) -- (6.5, -1);
																										
																										\draw [lineStyle, dashed] (6.5, -1) -- (7.5, -1);
																										\draw [lineStyle] (7.5, -1) -- (8.5, -1);
																										\draw [lineStyle,  ->] (8.5, -1) -- (9.5, -1);
																										\draw [lineStyle, dashed, ->] (9.5, -1) -- (10.5, -1);

																										\draw [myDot] (0,0) circle [radius=0.04];
																										\draw [myDot] (1,0) circle [radius=0.04];
																										\draw [myDot] (2,0) circle [radius=0.04];
																										\draw [myDot] (2.5, -1) circle [radius=0.04];
																										\draw [myDot] (3.5, -1) circle [radius=0.04];
																										\draw [myDot] (4.5, -1) circle [radius=0.04];
																										\draw [myDot] (5, -2) circle [radius=0.04];
																										\draw [myDot] (6, -2) circle [radius=0.04];
																										\draw [myDot] (6.5, -1) circle [radius=0.04];
																										\draw [myDot] (7.5, -1) circle [radius=0.04];
																										\draw [myDot] (8.5, -1) circle [radius=0.04];
																										\draw [myDot] (9.5, -1) circle [radius=0.04];
																										\draw [myDot] (10.5, -1) circle [radius=0.04];

																										\node[below left](call) at (2,0) {\tiny $call$};
																										
																										\node[left](proc-entry) at (2.3,-1) {\tiny \changeh{$proc$}};

																										\node[above] at (0,0) {\tiny $A_{0}$};
																										\node[below]() at (0,0) {\tiny $\changec{p_0 \omega_0}$};

																										\node[below right]() at (2,0) {\tiny $\changec{p_{x-t} \omega_{x-t}}$};
																										\node[above] at (2,0) {\tiny $A_{x-t}$};

																										\node[below]() at (2.5,-1) {\tiny $\changec{p_{x-(t-1)} \omega_{x-(t-1)}}$};
																										\node[above right] at (2.5,-1) {\tiny $A_{x-(t-1)}$};

																										\node[above]() at (4.5,-1) {\tiny $\changec{p_{x-y_1} \omega_{x-y_1}}$};
																										\node[below left](call) at (4.5,-1) {\tiny $call$};
																										\node[below right] at (4.5,-1) {\tiny $A_{x-y_1}$};
																										
																										\node[right](ret) at (6,-2) {\tiny $ret$};

																										\node[above]() at (6.5,-1) {\tiny $\changec{p_{x-y_2}, \omega_{x-y_2}}$};
																										\node[below left] at (6.5,-1) {\tiny $A_{x-y_2}$};

																										\node[below]() at (7.5,-1) {\tiny $\changec{p_{x-1} \omega_{x-1}}$};
																										\node[above] at (7.5,-1) {\tiny $A_{x-1}$};
																										
																										\node[below]() at (8.5,-1) {\tiny $\changec{p_{x} \omega_{x}}$};
																										\node[above] at (8.5,-1) {\tiny $A_{x}$};
																										
																										\node[below]() at (9.5,-1) {\tiny $\changec{p_{x+1} \omega_{x+1}}$};
																										\node[above] at (9.5,-1) {\tiny $A_{x+1}$};

																										\end{tikzpicture}
																										
																										\caption{Case of $X^c \phi' \in A_x$}
																										\label{fig:nextCaller2}
																									\end{figure*}

																								\end{myColorForRemovePages}

																								\else 

																								\ifdefined \NotNeedToReducePages	
																								
																								\begin{myColorForRemovePages}

																									\medskip
																									\noindent
																									\textbf{Encoding atoms to control locations.} Firstly, we need to ensure that $\mathcal{BP}_i$ has an accepting \changec{(local)} run
																									from $\changed{ \llparenthesis p_x, A_x \rrparenthesis\omega_x }$ \changec{iff $\changec{p_x \omega_x}$ satisfies $\phi$  (denoted $\changec{p_x \omega_x} \vDash \phi$)} for every $\phi \in A_x$. To ensure this, in rules $(\alpha_1)$, $(\alpha_2)$ and $(\alpha_3)$, the first class of conditions $(\beta_0)$ ensures that the tags
																									$\{ call, ret, int \}$ assigned to each configuration \change{of the run} are guessed correctly. The second class of conditions $(\beta_1)$ and $(\beta_2)$ \change{expresses} that for
																									every $e \in AP$, $(\changend{\uppi}, x) \vDash e$ iff $e \in \lambda(p_x)$, and the class of conditions $(\beta_3)$ \change{expresses} that $(\changend{\uppi}, x) \vDash X^g \phi'$ iff
																									$(\changend{\uppi}, x+1) \vDash \phi'$. Now, let us consider the \changenrp{most delicate case $\phi = X^{a} \phi' \in A_x$. There are two possibilities:}

																								\end{myColorForRemovePages}

																								\else
																								
																								\medskip
																								\noindent
																								\textbf{Encoding atoms to control locations.} Firstly, we need to ensure that $\mathcal{BP}_i$ has an accepting \changec{(local)} run
																								from $\changed{ \llparenthesis p_x, A_x \rrparenthesis\omega_x }$ \changec{iff $\changec{p_x \omega_x}$ satisfies $\phi$  (denoted $\changec{p_x \omega_x} \vDash \phi$)} for every $\phi \in A_x$. To ensure this, in rules $(\alpha_1)$, $(\alpha_2)$ and $(\alpha_3)$, the first class of conditions $(\beta_0)$ ensures that the tags
																								$\{ call, ret, int \}$ assigned to each configuration \change{of the run} are guessed correctly. The second class of conditions $(\beta_1)$ and $(\beta_2)$ \change{expresses} that for
																								every $e \in AP$, $(\changend{\uppi}, x) \vDash e$ iff $e \in \lambda(p_x)$, and the class of conditions $(\beta_3)$ \change{expresses} that $(\changend{\uppi}, x) \vDash X^g \phi'$ iff
																								$(\changend{\uppi}, x+1) \vDash \phi'$. Now, let us consider the \changenrp{most delicate case $\phi = X^{a} \phi' \in A_x$ . There are two possibilities:}
																								

																								\fi

																								\begin{figure*}
																									\centering
																									\begin{tikzpicture}[xscale=1.3, yscale=1.3]
																									\tikzset{lineStyle/.style={blue, ultra thick}};
																									\tikzset{myDot/.style={blue, fill = yellow, thick}};
																									\tikzset{myRectangleNode/.style={rectangle, thick, draw= black, below right, black}};
																									\draw [->, lineStyle, dashed] (0, 0) -- (1,0);
																									\draw [->, lineStyle] (1,0) -- (2,0) -- (2.5, -1) -- (3.5, -1) -- (4.5, -1);
																									\draw [->, lineStyle, dashed] (4.5, -1) -- (5.5, -1);
																									\draw [->, lineStyle] (5.5, -1) -- (6.5, -1) -- (7, 0) -- (8, 0);
																									\draw [->, lineStyle, dashed] (8, 0) -- (9, 0);
																									\draw [myDot] (0,0) circle [radius=0.04];
																									\draw [myDot] (1,0) circle [radius=0.04];
																									\draw [myDot] (2,0) circle [radius=0.04];
																									\draw [myDot] (2.5, -1) circle [radius=0.04];
																									\draw [myDot] (3.5, -1) circle [radius=0.04];
																									\draw [myDot] (4.5, -1) circle [radius=0.04];
																									\draw [myDot] (5.5, -1) circle [radius=0.04];
																									\draw [myDot] (6.5, -1) circle [radius=0.04];
																									\draw [myDot] (7,0) circle [radius=0.04];
																									\draw [myDot] (8,0) circle [radius=0.04];
																									\draw [myDot] (9,0) circle [radius=0.04];
																									\node[below left](call) at (2,0) {$call$};
																									
																									\node[above] at (0,0) {$A_{0}$};
																									\node[above right](AtomA_call) at (2,0) {$A_{x}$};
																									\node[above right]() at (2.5,-1) {$A_{x+1}$};
																									
																									\node[above left](proc-entry) at (2.3,-1) {\changeh{$proc$}};
																									\node[above left](ret) at (6.5,-1) {$ret$};
																									\node[below right](return-point) at (7,0) {\textit{return-point}};
																									\node[above left](AtomA0_pass) at (7,0) {$\llparenthesis \gamma'', A_{x} \rrparenthesis$};
																									\draw[dotted, red, thick, ->] (AtomA_call) .. controls(4,0.5) .. (AtomA0_pass);
																									\node[below right](encoded) at (3,0.3) {encoded $\&$ passed down};
																									\node[below]() at (0,0) {$p_0 \omega_0 $};
																									\node[below right]() at (2,0) {$\changec{p_x \omega_x}$};
																									\node[below]() at (2.5,-1) {$\changec{p_{x+1} \omega_{x+1}}$};
																									
																									\node[above ]() at (5.5,-1) {\changenta{$\changec{p_{y-2} \omega_{y-2}}$}};

																									\node[below]() at (6.5,-1) {$\changec{p_{y-1} \omega_{y-1}}$};
																									\node[above right]() at (7,0) {$p_y \omega_y$};
																									
																									\node[ below left]() at (6.9,0) {$A_{y}$};

																									\end{tikzpicture}
																									\vspace{-1em}
																									\caption{Case of $X^a \phi' \in A_x$ }
																									\vspace{-1.5em}
																									\label{fig:nextAbstract}
																								\end{figure*}
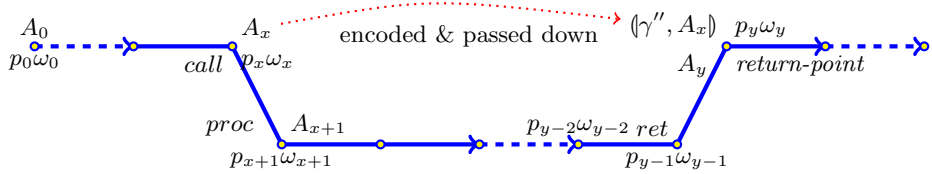

																								\ifdefined \NotNeedToReducePages \begin{itemize} \else \begin{itemize}[noitemsep,topsep=0pt] \fi
																										
																										\item
																										{ $\changec{p_x \omega_x}$ $\changec{\changend{\xRightarrow{}_{i}}}$ $\changec{p_{x+1} \omega_{x+1}} \changenta{\vartriangleright d_0}$ corresponds to a call statement.
																											Let us consider Figure \ref{fig:nextAbstract} to explain this case. Let $\changec{ p_y\omega_y }$ be the abstract-successor of $\changec{p_x \omega_x}$.
																											$(\changend{\uppi}, x) \vDash X^{a} \phi' $ iff $(\changend{\uppi}, y) \vDash \phi'$. Thus, we must have $\phi' \in A_{y}$.
																											This is ensured by rules \changenrpc{$(\alpha_1)$} and \changenrpc{$(\alpha_2)$}: rules $\changenrpc{(\alpha_1)}$ allow to record $X^a \phi'$ in the return point of the call, and rules $\changenrpc{(\alpha_2)}$ allow
																											to extract and validate $\phi'$ when the return-point is reached. In what follows, we show in more details how this works:
																											Let $\changec{p_x \gamma} \xrightarrow{call}_i \changec{p_{x+1}  \gamma' \gamma'' \vartriangleright d}$ be the rule associated with the transition $\changec{p_x \omega_x}$ $\changec{\changeh{\xRightarrow{}_{i}}}$ $\changec{p_{x+1} \omega_{x+1}} \changenta{\vartriangleright d_0}$, then we have $\omega_x = \gamma \omega'$ and $\omega_{x+1} = \gamma' \gamma'' \omega'$. Let $\changec{p_{y-1} \omega_{y-1}}$ $\changec{\changeh{\xRightarrow{}_{i}}}$ $\changec{p_{y} \omega_{y}} \changenta{\vartriangleright d_0}$ be the transition that corresponds to the $ret$ statement of this call.
																											Let then $\changec{p_{y-1} \beta} \xrightarrow{ret}_i \changec{p_y \epsilon} \vartriangleright d \in \Delta$ be the corresponding return rule.
																											Then, we have necessarily $\omega_{y-1} = \beta \gamma'' \omega'$, since as explained in Section \ref{sec:DPNsDefinition},
																											$\gamma''$ is the return address of the call. After applying this rule, $\omega_{y} = \gamma'' \omega'$.
																											In other words, $\gamma''$ will be the topmost stack symbol at the corresponding return point of the call.
																											So, in order to recover $\phi'$ in $A_{y}$, we proceed as follows:
																											At the call $\changec{p_x \gamma} \xrightarrow{call}_i \changec{p_{x+1}  \gamma' \gamma'' \vartriangleright d}$, we encode $A_x$ into $\gamma''$ by the rule
																											$(\alpha_1)$ stating that $\changec{\llparenthesis p_x, A_x \changeh{, l} \rrparenthesis \gamma} \changenrpd{\longrightarrow_i} \llparenthesis p_{x+1}, A_{x+1} \changeh{, l'}  \rrparenthesis \gamma' 
																											\llparenthesis \gamma'', A_x \changeh{, l} \rrparenthesis  \changenb{\vartriangleright d_0} \in \Delta'$. 
																											This allows to record $X^a \phi'$ in the corresponding return point of the stack.
																											After that, $\llparenthesis \gamma'', A_x \changeh{,l} \rrparenthesis $ will be the topmost stack symbol at the corresponding return-point of this call. 
																											At the return-point, the condition \changencd{$(\beta_6)$} in $(\alpha_{2.2})$ stating that $AbsNext(A_x, A_y)$ and the fact that $\phi = X^{a} \phi' \in A_x$
																											imply that $\phi' \in A_y$. 
																											

																											%
																											
																											\item{ \changenrpb{$\changec{p_x \omega_x}$ $\changec{\changend{\xRightarrow{}_{i}}}$ $\changec{p_{x+1} \omega_{x+1}} \changenta{\vartriangleright d_0}$ corresponds to a simple statement. Then, the abstract successor of $\changec{p_x \omega_x}$ is $\changec{p_{x+1} \omega_{x+1}}$.
																													$(\changend{\uppi}, x) \vDash X^{a} \phi' $ iff $(\changend{\uppi}, x+1) \vDash \phi'$. Thus, we must have $\phi' \in A_{x+1}$. This is ensured by condition $(\beta_4)$ in $(\alpha_3)$ stating that $AbsNext(A_x, A_{x+1}) = true$}
																											}

																										}
																									\end{itemize}

																									\fi

																									\medskip
																									\noindent
																									\textbf{The label \textit{l}.} \changeh{Now, let us explain how the label $l$ is used in the transition rules to ensure the correctness of the formulas. Note that our explanation above \changeh{makes} implicitly the assumption that along the run $\changend{\uppi}$, every call to a procedure $proc$ will eventually reach its corresponding return point, \changenb{i.e.}, the run $\changend{\uppi}$ will finally exit $proc$, then, we can encode formulas at the \textit{call} and validate \changeh{them} at its corresponding return-point. However, it might be the case that at a certain point in the procedure $proc$, there will be a loop, and $\changend{\uppi}$ never exits $proc$. To solve this problem, we annotate the control states by the label $l \in \{\changeh{exit}, \changeh{unexit}\}$ to determine whether $\changend{\uppi}$ can complete the execution of \changeh{the procedure} $\mathscr{P} (\changenrlangle{p \omega })$. In the following, we explain three cases \changeh{corresponding} to three kinds of statements: }

																									\ifdefined \NotNeedToReducePages \begin{itemize} \else \begin{itemize}[noitemsep,topsep=0pt] \fi
																											\item {\changeh{Let us consider Figure \ref{fig:nextAbstract}. $\changenrlangle{p_x \omega_x }$ $\changend{\xRightarrow{}_{i}}$ $\changenrlangle{p_{x+1} \omega_{x+1} } \changenta{\vartriangleright d_0}$ corresponds to a \textit{call} statement. \change{Note that} $\mathscr{P} (\changenrlangle{p_{x+1} \omega_{x+1} }) = proc$ in this case. There are two possibilities. If $proc$ terminates, then the call at $\changenrlangle{p_x \omega_x }$ will reach its \changeh{corresponding} return-point. In this case, $\changenrlangle{p_{x+1} \omega_{x+1} }$ is labelled by \textit{\changeh{exit}}. If $proc$ never terminates, then the call at $\changenrlangle{p_x \omega_x }$ will never reach its \change{corresponding} return-point. In this case, $\changenrlangle{p_{x+1} \omega_{x+1} }$ is labelled by \textit{\changeh{unexit}}. If $\changenrlangle{p_{x+1} \omega_{x+1} }$ is labelled by \textit{\changeh{exit}}, then $\changenrlangle{p_x \omega_x }$ can be labelled by \textit{\changeh{exit}} or \textit{\changeh{unexit}}. However, if $\changenrlangle{p_{x+1} \omega_{x+1} }$ is labelled by \textit{\changeh{unexit}}, then $\changenrlangle{p_x \omega_x }$ must be labelled by \textit{\changeh{unexit}}. This is ensured by the condition \changeh{($l' = \changeh{unexit}$ implies $l=\changeh{unexit})$} in the rule \changeb{$(\alpha_{1})$}. In addition, if $\changenrlangle{p_{x+1} \omega_{x+1} }$ is labelled by \textit{\changeh{unexit}}, then $\changenrlangle{p_x \omega_x }$ never reaches its corresponding return-point. Thus, $\changenrlangle{p_x \omega_x }$ does not satisfy any formula in the form $X^a \phi$. This is ensured by the condition \changeh{($l' = \changeh{unexit}$ implies $NexAbsForms(A) = \emptyset$)} in the rule \changeb{$(\alpha_{1})$}.}}

																											%

																											\item {\changeh{\changeh{Again,} let us consider Figure \ref{fig:nextAbstract}. $\changenrlangle{p_{y-1} \omega_{y-1} }$ $\changend{\xRightarrow{}_{i}}$ $\changenrlangle{p_y \omega_y } \changenta{\vartriangleright d_0}$ corresponds to a ret statement. At $\changenrlangle{p_{y-1} \omega_{y-1} }$, we are sure that $proc$ will terminate. In this case, $\changenrlangle{p_{y-1} \omega_{y-1} }$ must be always labelled by \textit{\changeh{exit}} and $\changenrlangle{p_y \omega_y }$ can be labelled by $\changeh{exit}$ or $\changeh{unexit}$. This is ensured by the rule \changenb{$(\alpha_{2.1})$}. Also, the abstract-successor of $\changenrlangle{p_{y-1} \omega_{y-1} }$ is $\bot$, then, $\changenrlangle{p_{y-1} \omega_{y-1} }$ does not satisfy \changeh{any formula} in the form $X^a \phi$. This is ensured by the condition ( $NexAbsForms(A)  = \emptyset$) in the rule \changeb{$(\alpha_{2.1})$}.}}

																											\ifdefined \NotNeedToReducePages	
																											
																											\begin{myColorForRemovePages}
																												
																												\item{\changeh{\changeh{Finally,} let us consider Figure \ref{fig:nextCaller2}. $\changenrlangle{p_x \omega_x }$ $\changend{\xRightarrow{}_{i}}$ $\changenrlangle{p_{x+1} \omega_{x+1} } \changenta{\vartriangleright d_0}$ corresponds to a \textit{simple} statement. Then, $\changenrlangle{p_x \omega_x }$ and $\changenrlangle{p_{x+1} \omega_{x+1} }$ are in the same procedure $proc$. Thus, the labels assigned to $\changenrlangle{p_x \omega_x }$ and $\changenrlangle{p_{x+1} \omega_{x+1} }$ should be the same. This is ensured by the transition rule \changeb{$(\alpha_{3})$} }}

																											\end{myColorForRemovePages}

																											\else
																											\item{\changeh{\changeh{Finally,} let us consider \changenrp{Figure \ref{fig:nextAbstract}. $\changenrlangle{p_{y-2} \omega_{y-2} }$ $\changend{\xRightarrow{}_{i}}$ $\changenrlangle{p_{y-1} \omega_{y-1} } \changenta{\vartriangleright d_0}$ corresponds to a \textit{simple} statement. Then, $\changenrlangle{p_{y-2} \omega_{y-2} }$ and $\changenrlangle{p_{y-1} \omega_{y-1} }$ are in the same procedure $proc$. Thus, the labels assigned to $\changenrlangle{p_{y-2} \omega_{y-2} }$ and $\changenrlangle{p_{y-1} \omega_{y-1} }$ should be the same. This is ensured by the transition rule \changeb{$(\alpha_{3})$} }}}
																											\fi

																										\end{itemize}

																										\ifdefined \NotNeedToReducePages	
																										
																										\begin{myColorForRemovePages}

																											\noindent
																											\textbf{The accepting conditions.} \changeh{The generalized B\"{u}chi accepting condition $F$ of $\mathcal{BP}_{i}$ consists of \change{three} families of accepting conditions \change{$F_1$, $F_2$ and $F_3$}. The first set $F_1$ guarantees that \changeh{an accepting run should go infinitely often through the label $\changeh{unexit}$}. Each set of $F_2$ ensures that the liveness requirement \change{$\phi_{2}$ in $\phi_{1} U^g \phi_{2}$} is eventually satisfied in $\mathcal{P}$.} Note that if $\phi_{1} U^g \phi_{2} \in A_i$, then, $(\changend{\uppi}, i) \vDash \phi_{1} U^g \phi_{2}$ iff $(\changend{\uppi}, i) \vDash \phi_2$ or ($(\changend{\uppi}, i) \vDash \phi_1$ and $(\changend{\uppi}, i) \vDash X^g (\phi_{1} U^g \phi_{2})$). Because $\phi_2$ should hold eventually, to avoid the case where the run of \changenb{$\mathcal{BP}_{i}$} always carries ($\phi_1$ and \change{$X^g (\phi_{1} U^g \phi_{2})$}) and never reaches $\phi_2$, we set $P \times F_{\phi_1 U^g \phi_2 } = P \times \{ A \in Atoms(f_i) \; | \;$ if $\phi_1 U^g \phi_2 \in A$ then $\phi_2 \in A \} $ as a set of B\"{u}chi generalized accepting condition. By this setting, the accepting run of $\mathcal{BP}_{i}$ will infinitely often visit some control locations in 
																											$P \times \{ A \in Atoms(f_i) \; | \;$ if $\phi_1 U^g \phi_2 \in A$ then $\phi_2 \in A \} $ which ensures that $\phi_2$ will eventually hold. \changeh{The idea behind the set $F_3$ is similar to the set $F_2$ except that the liveness requirement for a $U^a$-formula $\phi_1 U^a \phi_2$ is only required on the (unique) infinite abstract path (labelled by  $unexit$).} With respect to caller-until formulas, note that caller paths are always finite, 
																											so we do not need to consider this case in $F$. The liveness requirements of caller-until formulas are ensured by the condition 
																											$NexCallerForms(A) = \emptyset$. This requirement guarantees the liveness requirement $\phi_2$ in $\phi_1 U^c \phi_2$ eventually happens. Look at Figure \ref{fig:nextAbstract} for an illustration. Assume that $\phi_1 U^c \phi_2 \in A_{i+1}$, then, $(\changend{\uppi}, i+1) \vDash \phi_1 U^c \phi_2$ iff $(\changend{\uppi}, i) \vDash \phi_2$ or ($(\changend{\uppi}, i) \vDash \phi_1$ and $(\changend{\uppi}, i) \vDash X^c (\phi_1 U^c \phi_2)$). In other words, $\phi_1 U^c \phi_2 \in A_{i+1}$ iff $\phi_2 \in A_i$ or $(\phi_1 \in A_i$ and $X^c (\phi_1 U^c \phi_2) \in A_i)$. Since $\phi_2$ should eventually hold, $\phi_2$ should hold at $\changend{\uppi}(i)$ because $next^c_i = \bot$. 
																											To ensure this, we require that $NexCallerForms(A_{0}) = \emptyset$ which guarantees that $NexCallerForms(A_{i}) = \emptyset$. $NexCallerForms(A_{i}) = \emptyset$
																											ensures that the case $\phi_2 \in A_i$ occurs instead of $(\phi_1 \in A_i$ and $X^c (\phi_1 U^c \phi_2) \in A_i)$; which means that $(\changend{\uppi}, i) \vDash \phi_2$ and 
																											$\phi_2$ eventually holds. Notice that this requirement does not make any restriction for our algorithm:
																											given a CARET formula $f_i$, we can always obtain at least one atom $A$ such that $NexCallerForms(A) = \emptyset$. \changeh{Thus, we obtain the following lemma:}

																										\end{myColorForRemovePages}

																										\else
																										
																										\noindent
																										\textbf{The accepting conditions.} \changeh{The generalized B\"{u}chi accepting condition $F$ of $\mathcal{BP}_{i}$ consists of \change{three} families of accepting conditions \change{$F_1$, $F_2$ and $F_3$}. The first set $F_1$ guarantees that \changeh{an accepting run should go infinitely often through the label $\changeh{unexit}$}. Each set of $F_2$ ensures that the liveness requirement \change{$\phi_{2}$ in $\phi_{1} U^g \phi_{2}$} is eventually satisfied in $\mathcal{P}$.} \changeh{The idea behind the set $F_3$ is similar to the set $F_2$ except that the liveness requirement for a $U^a$-formula $\phi_1 U^a \phi_2$ is only required on the (unique) infinite abstract path (labelled by  $unexit$).} With respect to caller-until formulas, note that caller paths are always finite, 
																										so we do not need to consider this case in $F$. The liveness requirements of caller-until formulas are ensured by the condition 
																										$NexCallerForms(A) = \emptyset$ \changenta{since $\uppi(0)$ have no caller successors}.

																										\fi


																										\ifdefined \NotNeedToReducePages \begin{lemma} \else \begin{mylemma}\fi
																												\label{lemma:supportLemma}

																												\changeh{Given a DPDS $\mathcal{P}_i = (P, \Gamma, \Delta)$, and a CARET formula $f_i$, we can construct a GBDPDS $\mathcal{BP}_{i} = (P', \Gamma', \Delta', F) $ such that for every configuration $ p \omega  \in P_i \times \Gamma_i^*$, $p \omega \vDash f_i $ iff there exists an atom \changenk{$A \in Initial_i$ s.t.} \changenrpc{ $\mathcal{BP}_i$ has an accepting run from $ \llparenthesis p, A \changeh{,unexit} \rrparenthesis  \omega $}.} 
	
																												\ifdefined \NotNeedToReducePages \end{lemma} \else \end{mylemma}\fi

				
																									\notesnta{Can you please check this section? }
																										
																										\noindent 
\noindent
	\textbf{\changenta{Spawning new instances.}} Lemma \ref{lemma:supportLemma} guarantees that the problem of checking whether an instance of $\mathcal{P}_i$ \changenaa{starting from $p \omega$} satisfies $f_i$ can be reduced to the \changenaa{problem \changenaab{of checking} if  $\mathcal{BP}_i$ has an accepting run from $ \llparenthesis p, A \changeh{,unexit} \rrparenthesis  \omega $ where $A \in Initial_i$}. Now, we need to ensure the satisfiability on instances created dynamically. \changenta{Suppose that $\mathcal{P}_i$ \changenrpb{spawns} a new instance of $\mathcal{P}_j$ starting from $p_s \omega_s$, this means that we need to guarantee that $p_s \omega_s \vDash f_j$. Note that by applying Lemma \ref{lemma:supportLemma} for the DPDS $\mathcal{P}_j$, we get that  $p_s \omega_s \vDash f_j$ iff there exists an atom $A \in Initial_j$ \changenrpc{s.t.  $\mathcal{BP}_j$ has an accepting run from $ \llparenthesis p_s, A \changeh{,unexit} \rrparenthesis  \omega_s $}. Then, the requirement $p_s \omega_s \vDash f_j$ is ensured by the conditions   \changenb{$(\beta_6)$ in $(\alpha_1)$}, $(\beta_5)$ in $(\alpha_2)$ and $(\beta_6)$ in $(\alpha_3)$} stating that \changed{for every} $p  \gamma  \changenrpc{\xrightarrow{t}_i}  q  \omega  \vartriangleright d \in \Delta$ ($t \in \{call, ret, int\}$), we have $ \llparenthesis p, A \changenta{, l} \rrparenthesis \gamma  \changenrpd{\longrightarrow_i}\changed{ \llparenthesis q, A' \changenta{, l'} \rrparenthesis \omega} \vartriangleright d_0 \in \Delta'$ such that if $d = p_s \omega_s $, then, $d_0 = \llparenthesis p_s,A_0 \changenrpd{, unexit} \rrparenthesis \omega_s$ where $A_0 \in Initial_{j}$ (since $\delta(p_s \omega_s) = j$ in this case).

																										\notesh{Maybe we should show detailly how to compute $\mathcal{G}'$  }
																										
																										\medskip
																										\noindent
																										\change{Thus, we can show that:}
																										\begin{theorem}
																											\label{THEOREM:MAINTHEOREM}
																											\changez{Given a DPN $\mathcal{M} = \{ \mathcal{P}_1, ..., \mathcal{P}_n \}$, a single-indexed CARET formula $f = \bigwedge_{i=1}^n f_i$, we can compute a GBDPN $\mathcal{BM} = \{ \mathcal{BP}_1, ..., \mathcal{BP}_n\}$  such that a global configuration $\mathcal{G}$ of $\mathcal{M}$ satisfies $f$ iff  $\mathcal{G}' \in \mathcal{L} (\mathcal{BM} )$ where  $\mathcal{G}'$ is \changed{a global configuration of $\mathcal{BM}$ that corresponds to the configuration $\mathcal{G}$. }}

																										\end{theorem}

																										%
																										%
																										

																										\section{Single-indexed CARET model-checking for DPNs with regular valuations}
																										In this section, we consider the single-indexed CARET model-checking problem for DPNs with regular valuations, in which the set of configurations where an atomic proposition \change{is satisfied} is a regular language.
																										\begin{defn}
																											\label{def:regularSet}
																											Let $\mathcal{M} = \{ \mathcal{P}_1, ..., \mathcal{P}_n \}$ be a DPN. For every $i \in \{1..n\}$, a set of configurations of a pushdown process $\mathcal{P}_i= (P_i, \Delta_i, \Gamma_i)$ is \textit{regular} if it can be written as the union of sets of the form
																											$E_p$, where $p \in P_i$ and $E_p= \{(p,w)| w\in L_p \}$, where $L_p$ is a regular set over $\Gamma_i^*$.
																										\end{defn}
																										
																										\begin{defn}
																											\label{def:regularvaluation}
																											Let  $\mathcal{M} = \{ \mathcal{P}_1, ..., \mathcal{P}_n \}$ be a DPN. Let $AP$ be a finite set of atomic propositions. Let $\upnu : AP \rightarrow 2^{\bigcup_{i=1}^n P_i \times \Gamma_i^*}$ be a valuation. $\upnu$ is called \textit{regular} if for every $e \in AP$, $\upnu(e)$ is a regular set of configurations.
																										\end{defn}

																										Let $\upnu : AP \rightarrow 2^{\bigcup_{i=1}^n P_i \times \Gamma_i^*}$ be a regular valuation. We define $\lambda_\upnu : P\times\Gamma^* \rightarrow 2^{AP}$ such that
																										$\lambda_\upnu (\changec{p \omega })=\{e\in AP\mid p \omega \in\upnu (e)\} $.
																										Let $\changeh{\uppi} = \changec{ p_0 \omega_0 } \changec{ p_1\omega_1 }...$ be a local path of $\mathcal{P}_i$. We associate each configuration
																										$\changeh{ p_x \omega_x }$ of $\changeh{\uppi}$ with a tag
																										\changeh{$t_x$} in $\{ call, int, ret \}$ as presented in Section~\ref{sec:singleIndexedForDPNs}.
																										Let $f_i$ be a CARET formula over $AP$. The satisfiability relation \change{w.r.t.}  the regular valuation $\upnu$ is defined as follows:
																										$$\changeh{\uppi}\vDash_\upnu f_i \text{ iff } (\lambda_\upnu(\changec{ p_0 \omega_0} ),t_0)(\lambda_\upnu(\changec{ p_1 \omega_1 }),t_1) \cdots \vDash f_i$$
																										
																										\begin{theorem}
																											\label{prop:regularValuation}
																											\cite{DBLP:conf/aplas/SongT13}
																											Single-indexed LTL model-checking with regular valuations for DPNs can be reduced to standard LTL model checking for DPNs.
																										\end{theorem}

																										\ifdefined \NotNeedToReducePages	
																										
																										Given a DPN $\mathcal{M} = \{ \mathcal{P}_1, ..., \mathcal{P}_n \}$  and a regular valuation $\upnu : AP \rightarrow 2^{\bigcup_{i=1}^n P_i \times \Gamma_i^*}$, this result is based on translating every DPDS $\mathcal{P}_i$ ($i \in \{1..n\}$)
																										into a DPDS $\mathcal{P}' = \changenk{(P_i, \Gamma_i', \Delta_i' )}$ 
																										where the regular valuation requirements are encoded in $\Gamma_i'$.
																										The same reduction is still true for single-indexed CARET with regular
																										valuations. \changenrp{Due to lack of space we don't present the details here. We refer readers to \cite{DBLP:conf/aplas/SongT13}.
																											Therefore, given a single-indexed CARET formula with regular valuations and a DPN $\mathcal{M} = \{ \mathcal{P}_1, ..., \mathcal{P}_n \}$, we apply the reduction of \cite{DBLP:conf/aplas/SongT13} to obtain
																											a new DPN $\mathcal{M'} = \{ \mathcal{P'}_1, ..., \mathcal{P'}_n \}$ in which for every $i \in \{1..n\}$,  $\mathcal{P}_i' = \changenk{(P_i, \Gamma_i', \Delta_i')}$ s.t. the satisfiability of a CARET formula over $\mathcal{P}_i$ \change{w.r.t.}  the regular valuation $\upnu$
																											can be reduced to satisfiability of a CARET formula over $\mathcal{P}'_i$ with \changec{simple valuations}.} Therefore, we \change{can show that:}

																										\else

																										Given a DPN $\mathcal{M} = \{ \mathcal{P}_1, ..., \mathcal{P}_n \}$  and a regular valuation $\upnu : AP \rightarrow 2^{\bigcup_{i=1}^n P_i \times \Gamma_i^*}$, this result is based on translating every DPDS $\mathcal{P}_i$ ($i \in \{1..n\}$)
																										into a DPDS $\mathcal{P}' = \changenk{(P_i, \Gamma_i', \Delta_i' )}$ 
																										where the regular valuation requirements are encoded in $\Gamma_i'$.
																										The same reduction is still true for single-indexed CARET with regular
																										valuations. \changenta{For details about this reduction, we refer readers to \cite{DBLP:conf/aplas/SongT13}. We \change{can show that:}}
																										
																										\fi

																										\begin{theorem}
																											\label{theorem:regularValuationTheorem}
																											Single-indexed CARET model-checking with regular valuations for DPNs can be reduced to standard single-indexed CARET model checking for DPNs.
																										\end{theorem}

																										\section{DPNs Communicating via Locks}
																										
																										\notesh{Maybe we should add motivation for Lock-part. ONe reviewer of iFM doesn't why we need this?}
																										
																										\removed{Locks are a common technique to ensure mutual exclusion access between shared resources in multithreaded programs in reality. However, while DPNs allow to model a multithreaded program with procedure calls and thread creations, it can not be used to model concurrent programs where threads communicating via locks. Therefore, an model-checking algorithm for DPNs communicating via locks will allows us to use single-indexed CARET formulas to specify properties on such programs. In this section, we show how to  model-checking single-indexed CARET formulas against DPNs where the synchronization between processes are ensured through locks. }

																										
																										\changez{Dynamic Pushdown Network with Locks (L-DPNs) is a natural formalism for multithreaded programs \changed{communicating via locks} \cite{DBLP:conf/cav/LammichMW09,DBLP:journals/corr/SongT16}:}


																										\begin{defn}
																											A Dynamic Pushdown Network with Locks (\changez{L-DPN}) $\mathcal{M}$ is a set $\{\mathbb{L}, Act, \mathcal{P}_1, ..., \mathcal{P}_n \}$ where $\mathbb{L}$ is a set of locks, $Act = \{acq(l), rel(l), \change{\uptau} \; | \; l \in \mathbb{L} \}$ is a set of actions on locks s.t. $acq(l)$ (resp. $rel(l)$)  for $l \in \mathbb{L}$ \changez{represents an \textit{acquisition}} (resp. \textit{release}) of the lock $l$ and the action $\change{\uptau}$ \changez{describes} internal actions (neither acquire \changez{nor} release locks);  for every $1 \leq i \leq n$,   $ \mathcal{P}_i = (P_i, \Gamma_i, \Delta_i)$ is a Labelled Dynamic Pushdown System with Locks (L-DPDS), where $P_i$ is a finite set of control locations and $P_i \; \cap \; P_j = \emptyset $ for all $ j \neq i$, $\Gamma_i$ is a finite set of stack alphabets, and $\Delta_i$ is  a finite set of transitions rules. Rules of $\Delta_i$ are of the following form,  where $a \in Act$, $p, p_1 \in P_i,  \gamma \in\Gamma_i, \omega_1 \in \Gamma_i^*$,  \changez{$\changez{d} \in \{\changez{\Box}, p_s \omega_s \; | \; p_s \omega_s \in \bigcup_{1 \leq j \leq n}P_j \times \Gamma_j^* \} $}:

																											\ifdefined \NotNeedToReducePages \begin{itemize} \else \begin{itemize}[noitemsep,topsep=0pt] \fi

																													\item {$(r_1)$ $   p \gamma   \xrightarrow{(a, call)}_i  p_1 \gamma_1 \gamma_2 \vartriangleright \change{d} $}
																													\item { $(r_2)$ $   p \gamma   \xrightarrow{(a, ret)}_i  p_1 \epsilon \vartriangleright \change{d}  $}
																													
																													\item {$(r_3)$ $   p \gamma   \xrightarrow{(a, int)}_i  p_1 \omega  \vartriangleright \change{d}$}

																												\end{itemize}

																											\end{defn}
																											
																											\done{Explain L-DPNs rules here shortly}
																											

																											\ifdefined \NotNeedToReducePages	

																											\changez{Intuitively, a L-DPN is a DPN where processes communicate via locks.} \changenrp{The transition rules of L-DPNs are similar to DPNs where each rule is associated with an element in the set $\{call, ret, \changez{int}\}$ to denote whether the rule corresponds to a \textit{call}, \textit{ret} or a \textit{simple} statement (neither \textit{call} nor \textit{ret}).} The difference is that each transition rule of L-DPNs is assigned to one additional action \changenr{$a \in Act$}. Depending on the nature of the associated action $a$, each transition step of L-DPDSs include one additional operation on a given lock $l$. $acq(l)$ (resp. $rel(l)$)  \changez{represents an} \textit{acquisition} (resp. \textit{release}) of the lock $l$ and the action $\changez{\uptau}$ describe internal actions (neither \textit{acquire} \change{nor} \textit{release} locks).
																											
																											\else
																											
																											\changez{Intuitively, a L-DPN is a DPN where processes communicate via locks.} \change{The difference is that each transition rule of L-DPNs is assigned to one additional action \changenr{$a \in Act$}. Depending on the nature of the associated action $a$, each transition step of L-DPDSs include one additional operation on a given lock $l$. $acq(l)$ (resp. $rel(l)$)  \changez{represents an} \textit{acquisition} (resp. \textit{release}) of the lock $l$ and the action $\changez{\uptau}$ describe internal actions (neither \textit{acquire} \change{nor} \textit{release} locks).}
																											\fi

																											A \textit{local configuration} of an instance of a \textit{L-DPDS} $\mathcal{P}_i$ is a tuple $(p \omega, L)$ where $p \in P_i$ is the control location, $\omega \in \Gamma_i^*$ is the stack content and $L \subseteq \mathbb{L}$ is a set of locks owned \change{by} the instance. A \textit{global configuration} of $\mathcal{M}$ is a multiset over $ \bigcup_{1 \leq i \leq n}P_i \times \Gamma_i^* \times 2^{\mathbb{L}}$, in which $(p \omega, L) \in P_i \times \Gamma_i^*  \times 2^{\mathbb{L}}$ represents the \textit{local configuration} of an instance of a pushdown process $\mathcal{P}_i$  which is running in \changec{the network}.

																											\medskip
																											\noindent
																											A L-DPDS $\mathcal{P}_i$ defines a transition relation $\changeh{\xRightarrow{}_{i}}$ as follows where $t \in \{call, ret, int\}$:
																											\ifdefined \NotNeedToReducePages \begin{itemize} \else \begin{itemize}[noitemsep,topsep=0pt] \fi
																													\item {if $p \gamma \xrightarrow{(\change{\uptau}, t)}_i p_1 \omega_1 \change{\vartriangleright d} $ then $(p \gamma \omega, L) \changeh{\xRightarrow{}_{i}} (p_1 \omega_1 \omega, L) \change{\vartriangleright D_0}$ \changez{where $D_0 = \emptyset$ if $d = \Box$, $D_0 = \{ (p_s \omega_s, \emptyset) \}$ if $d = p_s \omega_s$} for every $\omega \in \Gamma_i^*$, $L \subseteq \mathbb{L}$ }

																													\item {if $p \gamma \xrightarrow{(acq(l), t)}_i p_1 \omega_1 \change{\vartriangleright d} $ then $(p \gamma \omega, L ) \changeh{\xRightarrow{}_{i}} (p_1 \omega_1 \omega, \changez{L \cup \{l\}}) \change{\vartriangleright D_0}$  \changez{where $D_0 = \emptyset$ if $d = \Box$, $D_0 = \{ (p_s \omega_s, \emptyset) \}$ if $d = p_s \omega_s$} for every $\omega \in \Gamma_i^*$, $L \subseteq \mathbb{L}$. This expresses that the current instance can move from $(p \gamma \omega, L)$ to $(p_1 \omega_1 \omega, \changez{L \cup \{l\}})$. This \changez{ensures} that the current instance \changez{owns} the lock $l$ after the action $acq(l)$.}
																													
																													\item {if $p \gamma \xrightarrow{(rel(l), t)}_i p_1 \omega_1 \change{\vartriangleright d} $ then $(p \gamma \omega, L ) \changeh{\xRightarrow{}_{i}} (p_1 \omega_1 \omega, \changez{L \setminus \{l\}}) \change{\vartriangleright D_0}$  \changez{where $D_0 = \emptyset$ if $d = \Box$, $D_0 = \{ (p_s \omega_s, \emptyset) \}$ if $d = p_s \omega_s$} for every $\omega \in \Gamma_i^*$, $L \subseteq \mathbb{L}$. This means that the current instance can move from $(p \gamma \omega, L)$ to $(p_1 \omega_1 \omega, \changez{L \setminus \{l\}})$. This \changec{ensures} that the current instance \changez{releases} the lock $l$ after the action $rel(l)$.}


																												\end{itemize}

																												\notes{About the question why $\changeh{\xRightarrow{}_{i}}$ of L-DPNs does not have reflexive and transitive closure, it is because we do not need it here. DPNs need it because the way it check membership is in a recursive manner, it need to record the set of DCLICs, I think we can remove the reflexive and transitive closure in DPNs for consistent with Lock-DPNs. We don't care it, it is used only when checking membership. WHat do you think????}
																												
																												\changez{Roughly speaking, if \changenr{$d = p_s \omega_s$}, then the current instance not only does local move but also creates a new instance of the pushdown process $\mathcal{P}_j$ starting at $(p_s \omega_s, \emptyset)$. Note that we suppose that the new instance holds no locks when it is created.}

																												A \textit{local run} of an instance of a L-DPDS $\mathcal{P}_i$ starting at a local configuration $c_0$ is a sequence  $c_0c_1...$ s.t. for every $j \geq 0$, $c_j \in P_i \times \Gamma_i^* \times 2^{\mathbb{L}}$ \changed{is a local configuration of  $\mathcal{P}_i$, $c_j \changeh{\xRightarrow{}_{i}}  c_{j+1} \vartriangleright D_0$}. The definition of \textit{global run}  of a L-DPNs $\mathcal{M}$ is similar to the one for DPNs.

																												
																												\noindent
																												\textbf{Nested Lock Access.} In this work, we suppose that in all local runs, the locks are accessed in a well-nested and no-reentrant manner, \change{i.e.} \changec{a} local \changed{run} can only release the latest lock \changez{it acquired that is not released yet}. \changez{Indeed, if we allow arbitrary locks, then reachability becomes undecidable \cite{DBLP:conf/lics/KahlonG06}. }

																												
																												\begin{theorem}
																													\label{prop:regularValuation}
																													\changenrpc{\cite{DBLP:journals/corr/SongT16}Single-indexed LTL model-checking for L-DPNs can be reduced to single-indexed LTL model checking for DPNs.}
																												\end{theorem}


																												\ifdefined \NotNeedToReducePages	
																												
																												\changeh{Given a L-DPN $\mathcal{M} = \{ \mathcal{P}_1, ..., \mathcal{P}_n \}$, this result is based on translating every $\mathcal{P}_i$ ($i \in \{1..n\}$) into a DPDS  $\mathcal{P}'_i =(P'_i, \Gamma_i, \Delta'_i)$ s.t. $\mathcal{P}'_i$ is a kind of product between the DPDS $\mathcal{P}_i$ and the acquisition structure, where an acquisition structure (encoded in control locations of $\mathcal{P}'_i$) stores information about how locks are used such as the number of held locks, the order of acquisition and release of locks. We can compute a DPN $\mathcal{M}' = \{ \mathcal{P}'_1, ..., \mathcal{P}'_n \}$ s.t. the global runs of $\mathcal{M}'$ mimic the global runs of $\mathcal{M}$ and the acquisition structures reflect the lock usages. Thus, the global runs of $\mathcal{M}'$ correspond to global runs of  $\mathcal{M}$ in which the locks are accessed in a nested manner. The same reduction is still true for single-indexed CARET formulas. For details of this reduction, we refer readers to \cite{DBLP:journals/corr/SongT16}. \changenrp{Therefore, given a single-indexed CARET formula and a L-DPN $\mathcal{M} = \{ \mathcal{P}_1, ..., \mathcal{P}_n \}$, we apply the reduction of \cite{DBLP:journals/corr/SongT16} to obtain a DPN $\mathcal{M'} = \{ \mathcal{P'}_1, ..., \mathcal{P'}_n \}$ s.t. the satisfiability of a CARET formula over $\mathcal{P}_i$ can be reduced to  satisfiability of a CARET formula over $\mathcal{P}'_i$} . Therefore, we \change{can show that:}}

																												\else
																												
																												\changenta{Given a L-DPN \changenrpd{$\mathcal{M} = \{\mathbb{L}, Act, \mathcal{P}_1, ..., \mathcal{P}_n \}$}, this result is based on translating every $\mathcal{P}_i$ ($i \in \{1..n\}$) into a DPDS  $\mathcal{P}'_i =(P'_i, \Gamma_i, \Delta'_i)$ s.t. $\mathcal{P}'_i$ is a kind of product between the DPDS $\mathcal{P}_i$ and the acquisition structure, where an acquisition structure (encoded in control locations of $\mathcal{P}'_i$) stores information about how locks are used such as the number of held locks, the order of acquisition and release of locks. We can compute a DPN $\mathcal{M}' = \{ \mathcal{P}'_1, ..., \mathcal{P}'_n \}$ s.t. the global runs of $\mathcal{M}'$ mimic the global runs of $\mathcal{M}$ and the acquisition structures reflect the lock usages. Thus, the global runs of $\mathcal{M}'$ correspond to global runs of  $\mathcal{M}$ in which the locks are accessed in a nested manner. The same reduction is still true for single-indexed CARET formulas. For details of this reduction, we refer readers to \cite{DBLP:journals/corr/SongT16}. We \change{can show that:}}
																												
																												\fi

																												\begin{theorem}
																													\label{theorem:lockDPNsTheorem}
																													Single-indexed CARET model-checking for L-DPNs can be reduced to single-indexed CARET model checking for DPNs.
																												\end{theorem}

																												\bibliographystyle{plain} 

																												\ifdefined \UseFullVersion
																												
																												\input{dpnProofs.tex}

																												\fi
																												
																											\end{document}